\documentclass[twocolumn,letterpaper, oldfontcommands]{IEEEAerospaceCLS}
\usepackage{cite}
\usepackage{siunitx}
\usepackage{float}

\usepackage{amsmath,amssymb,amsfonts, amstext}
\usepackage{algorithmic}
\usepackage{graphicx}
\usepackage{textcomp}
\usepackage{xcolor}

\usepackage{hyperref}

\usepackage{caption}
\usepackage{subcaption}
\usepackage{tabstackengine}
\stackMath
\usepackage{multirow}

\usepackage[capitalize, noabbrev]{cleveref}

\usepackage{array}   
\newcolumntype{L}{>{$}l<{$}} 
\newcolumntype{R}{>{$}r<{$}} 

\begin{document}

\title{High-precision Hardware Oscillators Ensemble for GNSS Attack Detection}

\author{Marco Spanghero\\
\textit{Networked Systems Security Group} \\
\textit{KTH Royal Institute of Technology}\\
Stockholm, Sweden \\
marcosp@kth.se
\and
Panos Papadimitratos \\
\textit{Networked Systems Security Group} \\
\textit{KTH Royal Institute of Technology}\\
Stockholm, Sweden \\
papadim@kth.se
}
\maketitle

\thispagestyle{plain}
\pagestyle{plain}

\maketitle

\thispagestyle{plain}
\pagestyle{plain}

\begin{abstract}

A wide gamut of important applications rely on global navigation satellite systems (GNSS) for precise time and positioning. Attackers dictating the GNSS receiver position and time solution are a significant risk, especially due to the inherent vulnerability of GNSS systems. A first line of defense, for a large number of receivers, is to rely on additional information obtained through the rich connectivity of GNSS enabled platforms. Network time can be used for direct validation of the GNSS receiver time; but this depends on network availability. To allow attack detection even when there are prolonged network disconnections,  we present a method based on on-board ensemble of reference clocks.  This allows the receiver to detect sophisticated attacks affecting the GNSS time solution, independently of the specific attack methodology. Results obtained with Chip-Scale Oven Compensated Oscillators (CS-OCXO) are promising and demonstrate the potential of embedded ensembles of reference clocks, detecting attacks causing modifications of the receiver time offset as low as \SI{0.3}{\micro \second}, with half the detection latency compared to related literature.
\end{abstract}

\tableofcontents

 \section{Introduction}
 
Global navigation satellite systems (GNSS) receivers provide time and position information to a wide gamut of applications, from mainstream smartphones to power systems. Due to the inherent vulnerability of GNSS, most notably the absence of publicly available authenticated signals \cite{walker2015galileo, anderson2017chips}, currently, adversaries can tamper with the GNSS-derived information \cite{psiaki2016gnss, schmidtSurveyAnalysisGNSS2016}.

It is possible to build and deploy an adversarial GPS signal transmitter steering the receiver Position, Velocity and Time (PVT) solution even with a limited budget \cite{Humphreys2008, borowski2012detecting, tippenhauer2011requirements, Humphreys2012a, PapadimitratosJa:C:2008}. 

Modern GNSS receivers are often integrated in (highly) connected mobile devices, so that deviations in the PVT solution can be detected by cross-checking the GNSS-provided information against that provided by network-based services. Specifically for time-based detection of GNSS attacks, the so-called Time Test can be leveraged \cite{papadimMilcom2008}. In this case, various secure network time providers (NTS, PTP) can be leveraged \cite{spangheroGNSS20, kzmsppPLANS2020} to compare the time solution provided by the GNSS receiver with a trusted time reference. However, a diversified attacker can manipulate GNSS signals to induce a false PVT solution while preventing the receiver from connecting to the network altogether or, selectively, to network time providers. 

Without (guaranteed) network connectivity, time-based detection can rely on the on-board capabilities of an enhanced GNSS receiver, notably its on-board clock \cite{PapadimitratosJa:C:2008}. Adversary-induced degradation of the time solution, manifesting itself as variations in the GNSS time rate and skewing, can be captured by taking advantage of on-board oscillators \cite{Arafin2016DetectingOscillators, Arafin2017, marnach2013detecting}. Intuitively, the progression of the GNSS receiver time should match the one provided by the local oscillator, if it is sufficiently stable during the period of observation \cite{Hwang2014, Spanghero2021a}. 
Chip-scale atomic clocks (CSAC) can address this by providing a long-term stable time reference, outperforming most solutions based on traditional oscillators. On the other hand, such devices are unlikely to be largely adopted due to the high cost and implementation complexity, limiting their scope to few applications \cite{Krawinkel2015, Gaggero2008}. Methods based on the fusion of multiple reference oscillators show significantly improved stability compared to a single embedded clock \cite{Levine1999c}.  This leads to the use of ensembles of oscillators as a cost-effective approach towards time-based detection of attacks. 

In this paper, we revisit the so-called Time Test approach that compares the GNSS time to the local time reference. We investigate how to obtain low-cost and stable on-board time reference that allows effectively detecting GNSS attacks. We develop a method based on multiple hardware oscillators deployed on Field Programmable Gate Array (FPGA), allowing high-accuracy measurement of the GNSS clock phase and frequency offset. An attack detection is performed by a statistical test on the expected GNSS receiver clock drift and offset. We extend existing methods based on a single clock, relying on a stochastic filter to provide a virtual clock combining multiple local oscillators.
Our solution is agnostic of the specific type of GNSS constellation or adversarial behavior. Scenarios for static or mobile receiver and attackers are evaluated, showing effective detection for message and code aligned spoofers. 
More specifically, relying on multiple oven-compensated chip-scale oscillators (OCXO) we get encouraging results: we detect relative deviations of the GNSS time as low as \SI{0.3}{\micro\second}, thus detecting an attack before it reaches its target time offset. 
Our system provides a stand-alone, low-cost time assurance device enabling high stability, effective detection. Moreover, it can be a building block for solutions that also have network-obtained accurate time.

The rest of this paper is structured as follows: \cref{sec:related-work} discusses relevant related work, \cref{sec:system-attacker-model} presents the system fundamentals and the adversary model. \cref{sec:proposed-approach,sec:implementation-experimental} present the proposed method used for GNSS attack detection, its implementation for ensemble-based GNSS misbehavior detection and the experimental setup. \cref{sec:evaluation} provides an evaluation of the proposed method and \cref{sec:conclusion} concludes with a discussion of possible future developments.

\section{Related work}
\label{sec:related-work}
The usage of clock-based countermeasures to detect GNSS attackers has been explored in literature as clock properties like offset and frequency stability can be used to identify misbehavior in the GNSS receiver, potentially caused by an adversary. Detection methods with time-based tests are effective against detecting attacks and are often agnostic of the specific attack strategies that affect the time offset computed by the victim receiver \cite{PapadimitratosJa:C:2008, papadimMilcom2008}. Beyond the aforementioned early works on time-based detection for GNSS attacks, we discuss next the most closely related recent works, including notably approaches that rely on local time references or an ensemble of clocks. Nonetheless, we do not discuss time-based detection approaches that rely on network-obtained time information \cite{kzmsppPLANS2020, spangheroGNSS20}, which are a powerful tool, but are dependant on the quality of the connection, and the level of sophistication of the attacker able to tamper with (access to) the remote time server.

In \cite{marnach2013detecting}, meaconing attacks are detected by means of a statistical test on the receiver clock offset. While having high resolution (\SI{80}{\nano \second}), the approach has a series of limitations. Most importantly, the performance is highly dependant on the receiver's clock temperature, needing to be practically constant over long periods of time. This is hardly possible in realistic scenarios, but the shortcoming can be partially resolved with higher performance clocks (i.e., oven and dual-oven compensated oscillators). Additionally, the threshold for detection requires to be set based on the estimated minimum duration of the delay introduced by the meaconer, requiring a per-case configuration.

In \cite{shangClockSingleSignal}, a technique based on repeated measurements of a single GNSS signal is presented, identifying spoofing signals based on the characteristic clock drift of the adversarial transmitter. The proposed countermeasure requires that the legitimate receiver is equipped with a low drift clock, e.g., an atomic clock or an oven compensated oscillator. The system detects effectively spoofing events and other attacker characteristics, such as the spoofer clock accuracy, but it requires a coarse knowledge of the legitimate receiver position and was demonstrated to be effective only for static receivers.

In \cite{Hwang2014}, a method that provides autonomous signal authentication based on the receiver clock stability is presented, allowing the receiver to distinguish between legitimate and adversarial signals. On the other hand, this countermeasure relies on the receiver's movement (even if minimal, i.e, oscillating antennas) to isolate counterfeit signals. As a result, it cannot safeguard static receivers. Additionally, \cite{Hwang2014} and \cite{shangClockSingleSignal} require access to low level signal properties that often are not available in consumer grade receivers.

To provide local time reference based validation, \cite{Arafin2016DetectingOscillators, Arafin2017} propose a method based on Kalman filters to track the clock and frequency offset between the GNSS receiver oscillator and an off-the-shelf, inexpensive temperature compensated oscillator (TCXO). The performance is measured based on the attack proposed in \cite{Shepard2012c}, where GNSS signals are progressively delayed, forcing the GNSS receiver to deviate from the ground truth time. While the authors consider the possibility of integrating various clocks to compensate for individual behaviour, improvements obtained from multiple clocks are not fully explored. Chip scale oven compensated oscillators (CS-OCXO) (and even more the combination of multiple ones), provide better performance, compared to lower grade TCXO, allowing for lower detection threshold and faster detection.

To take advantage of multiple on-board oscillators, the literature on high precision synchronization provides multiple approaches to achieve precision time-keeping with an ensemble of lesser quality clocks. In \cite{JonesTryonTimeSeries}, a continuous time series model for unequally spaced atomic clocks time data is provided and the application of Kalman filters for clock parameter tracking is also described in \cite{Greenhall2001,Greenhall2011b}, to generate computationally efficient time scales. This allows obtaining a virtual clock whose performances is better than the performance of any single clock in the ensemble. Originally developed for atomic clocks, clock ensembling can be used for commercial off the shelf (COTS) oven-compensated oscillators too \cite{Levine1999c}. Although the approach of creating clock ensembles is well-established in theory and various simulation results exist (e.g., for atomic time keeping), there are few results with implementation in embedded devices and often demonstrations are limited to specialized contexts (\cite{Zenzinger2012, Peiffer2016}), especially with low-cost oscillators. 

Current literature on clock ensembling is focused on increasing the overall time-keeping performance, but, practically, limited results towards applying clock ensembles in the context of GNSS attack detection are available. The implementation of an enhanced receiver comprising an embedded clock ensemble could be beneficial to achieve attack detection at a fraction of the cost and space/power footprint compared to existing solutions.

\section{System and Adversary model}
\label{sec:system-attacker-model}
After a brief description of the GNSS basics, we provide here background on clock characteristics and performance to facilitate the presentation of our approach. Then, we discuss relevant aspects of the adversarial behavior.

\subsection{GNSS basics}

GNSS systems rely on difference of time of arrival to estimate the position of the receiving antenna relative to a constellation of satellites. The receiver uses a local oscillator to measure the satellite-receiver GNSS signal propagation time, which corresponds, essentially, to a satellite-receiver distance estimate termed pseudorange. Given a vector of pesudoranges, $\rho$, a satellite observation matrix, $H$, and a vector of noise processes, $W$, the position and time offset solution ($s = [x(t), y(t), z(t), \delta_t(t)]$) \footnote{Velocity is obtained by combination of the calculated Dopper shift of the satellite signals and receiver motion} is obtained at the GNSS receiver by solving for $s$ the following linear system:
 \begin{equation}
     \rho = H*s + W
 \end{equation}

While atomic clocks on board the satellites are kept synchronized by a complex ground control infrastructure, the receivers typically rely on a cost-effective crystal oscillator. As errors in the receiver clock synchronization cause loss of accuracy for the PVT solution, constant tuning is needed due to poor stability characteristics of the receiver's clock. Thus, the GNSS receiver estimates the local clock bias alongside with its location coordinates; at least four satellites are required at any time to provide a valid PVT solution.

Using the time offset correction, $\delta_t$, obtained from the constellation, a GNSS receiver can constantly correct the local time to the one provided by the GNSS constellation:
\begin{equation}
    t_{corrected} = t_{local} + \delta_t
\end{equation}
Practically, the achievable timescale accuracy at the receiver depends on the receiver type and the surrounding environment (i.e., static or dynamic receiver, number of satellites in view). Typically, in a benign setting an accuracy of \SI{30}{\nano\second} or better can be achieved with commercial grade GNSS receivers \cite{USNOAccuracy, ESAAccuracy}.

\subsection{Clock model}



\label{section:clock-model}
The simplest definition of a clock is an oscillator connected to a counter and system time is defined as progression of the value of the counter. An observer would obtain the current time by reading the counter at specific interval. As the counter is driven by a reference oscillator, the error in the time information is due to the change in the frequency rate of the oscillator itself \cite{Cantor1999}.

In \cite{JonesTryonTimeSeries,Greenhall2001} a comprehensive model for atomic clocks is given. We adapt this model, based on \cite{Kim2012}, to low-precision oscillators with a clock state vector defined as $x(t_n) = [\theta(t_n), \gamma(t_n), D(t_n)]$, where $\theta$, $\gamma$ and $D$ indicate the clock offset, the frequency offset and the frequency drift at epoch $t_n$. The offsets $\theta, \gamma$ are residuals between the clock state and an ideal clock at the same epoch. The clock produces periodic events (pulses) at a rate defined by the clock frequency. The difference between the rate of the clock and the rate of an ideal reference is defined as skew and can be obtained by comparing the increments in each clock at chosen sampling instants, where $\theta(t)$ is the offset between the ideal clock and the clock under test.


Specifically, the clock offset can be written as a linear combination of offset, rate and drift as in \cref{eqn:clock-linear}, where $\theta(0)$ is the initial offset and $w(t)$ represents the non-deterministic random variations of the clock:

\begin{equation}
    \theta(t) = \theta_0 + \gamma \cdot t + \frac{1}{2} \cdot D \cdot t^2 + w(t)
    \label{eqn:clock-linear}
\end{equation}

Due to non-deterministic errors caused by manufacturing variations, each clock is unique. The stability of each clock has individual characteristics due to the fact that frequency tuning and stability are unique to each device \cite{Kohno2005}.
Experimental evidence of this fact is shown in Figure \ref{fig:hadamard-independent}, where each of the three clocks in our system is measured over an extended period of time against a high-quality reference. 

\subsubsection{Realization of hardware clock ensembles}
Clock ensembles are an option to achieve extended stability: the intuition behind clock ensembles is that multiple oscillators can be combined to improve over their stability, compensating for errors. This approach relies on considering each clock as a stochastic process characterized by a state vector, $x(t)$,  and an estimator tracking individual clocks and estimating corrections \cite{JonesTryonTimeSeries}.

\begin{figure}
    \centering
  \includegraphics[width=0.8\linewidth]{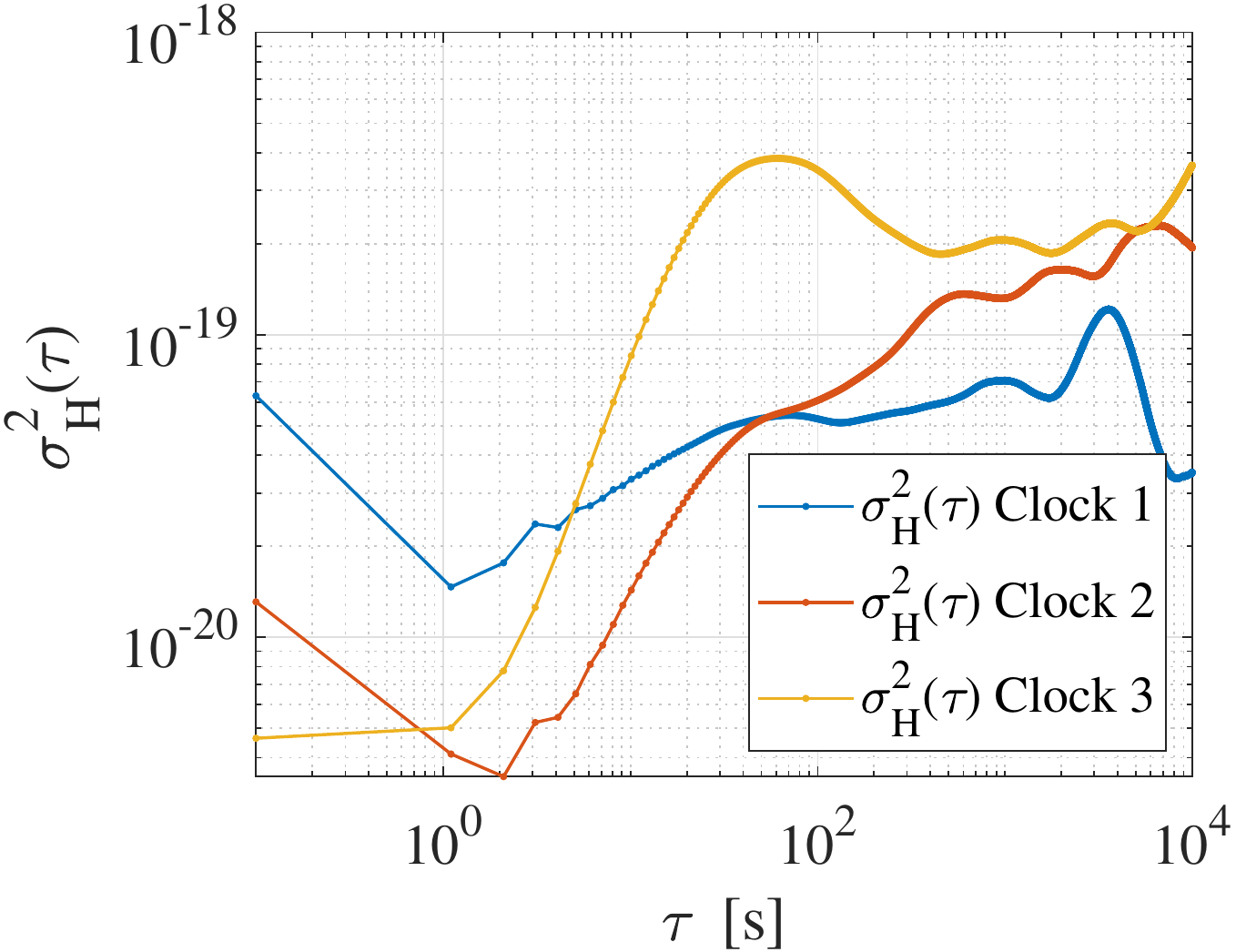}
  \caption{Nominal Hadamard variance of three independent OCXO conditioned using a Phase Locked Loop (PLL)}
  \label{fig:hadamard-independent}
\end{figure}

Specifically, we rely on the methods presented in \cite{Brown1991b, Greenhall2006}, using a Kalman filter to estimate the differential states of the clocks and produce corrections for the individual members of the ensemble. Given the state vector $x(t)$, the clock is modeled with the following stochastic differential equation, with step $\tau$:
\begin{equation}
    \begin{cases} 
    \frac{d\theta}{d\tau} = \gamma + w_1 \\
    \frac{d\gamma}{d\tau} = D + w_2 \\
    \frac{dD}{d\tau} = w_3
    \end{cases} 
    \label{eq:clock-model-diff}
\end{equation}
where $w_1, w_2, w_3$ represents the zero-mean Gaussian noise associated with each process characterized by spectral densities $q_1,q_2,q_3$, obtained from the Hadamard variance \cite{JonesTryonTimeSeries}.
Figure \ref{fig:hadamard-independent} shows the Hadamard variance for three independent clocks that are used in our system. The Hadamard variance is calculated as:

\begin{equation}
    \sigma_H^2(\tau) = \frac{1}{6(M-2)}\sum_{i=1}^{M-2}[y_{i+2} - 2y_{i+1} + y_i]^2
\end{equation}

where $y_i$ is the i-th of M fractional frequency values at averaging time $\tau$. The Hadamard variances are measured over a period of \SI{1e4}{\second} using an Agilent 53131A universal counter at a sampling rate of \SI{1}{\hertz}.

Tracking the state of multiple clocks in an ensemble poses various challenges. Each clock model is defined with respect to an \textit{ideal} clock: for this reason we cannot directly measure the clock states, as ideal time sources are, as a matter of fact, unavailable. 
Instead, we measure the differences between each clock and one clock within the ensemble that shows the best stability (master clock relativization \cite{Brown1991b}). In benign conditions (absence of attackers), the GNSS-derived clock achieves the best stability, thus it is regarded as master clock. Because we can only measure clock differences, the system is not fully observable: this causes the covariance estimation for the promoted master clock to grow unbounded \cite{Brown1991b}. While this is not problematic for the Kalman filter itself, it could cause numerical instability and overflow issues. This issue can be solve with the techniques described in \cite{Brown1991b} and \cite{Greenhall2011b}. 

\subsection{Attacker model}
\begin{figure}[h]
  \centering
  \includegraphics[width=0.7\linewidth]{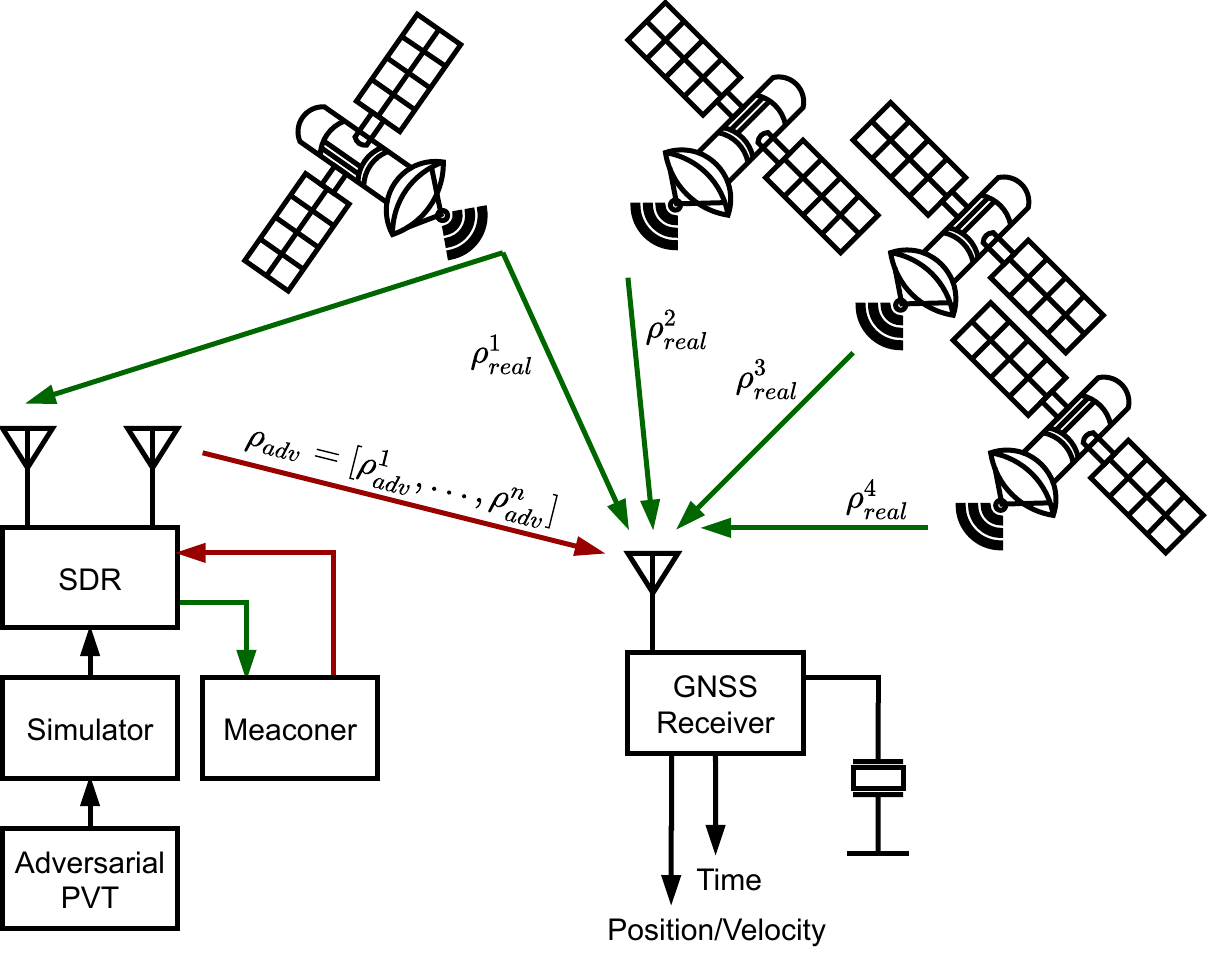}
  \caption{GNSS attacker: depending on the type of attack, the adversary can replay legitimate GNSS signals (meaconer) or create artificial ones (simulator).}
  \label{fig:attacker-diagram}
\end{figure}

An attacker can control the victim by carefully crafting or replaying GNSS signals to produce an erroneous PVT solution at the victim receiver, as illustrated in \cref{fig:attacker-diagram}. We are largely agnostic to the adversarial method for attacking the receiver, in order to control its PVT solution. We classify adversaries in a simple manner, as naive and advanced, based on their knowledge (or lack thereof) of the victim receiver PVT state and legittimate GNSS signals. 

\textit{Naive} attackers modify the Position/Time information without knowledge of the victim's PVT solution at the beginning of the attack. These attacks produce sharp discontinuities in the PVT solution: thus, monitoring for abrupt PVT changes can reveal such attacks. GNSS signal simulation or replay/meaconing attacks that are not synchronized with the current GNSS signal cause such effects. Moreover, it is typically required to first jam before succeeding in having the victim lock on the adversarial signals. For static victim receivers, attackers with knowledge of the victim's position can produce consistent adversarial signals, but this still would cause abrupt changes in the victim receiver's clock offset at the beginning of the attack due to the lack of synchronization.

On the contrary, \textit{advanced} attackers, having knowledge of the legitimate GNSS signals and PVT at the victim receiver, can produce a more subtle effect synchronizing the code phase and navigation message of the adversarial signals to the legitimate ones. Such adversaries do not introduce sudden variations in the victim receiver PVT solution but slowly 'drag' the victim GNSS information away towards their objective. Such adversaries are more challenging to detect, as the variation they introduce can be subtle enough to avoid detection. Most modern receivers cannot detect such slow variations and few (not widespread nor comprehensive) cases of embedded detection techniques exist in commercially available receivers (i.e., as described in \cite{ZEDF9PInterfaceDescription}).

A specific case of advanced meaconing consists of replaying GNSS signals with variable delay \cite{Blum2019}. Such attack preserves all the characteristics of the real signal and can target cryptographically secure signals. For Secure Code Estimation and Replay attackers (SCER), it is unfeasible to maintain a low clock offset and a low position offset at the beginning of an attack if the delay between the spoofed signals and the real signals is kept to a minimum, as either quantity will experience a significant discontinuity, giving means for detection \cite{humphreys2013detection}. 

Independently of the attacker's sophistication, the attack complexity also depends on the mobility of the victim receiver: static receivers can be overtaken by an attacker in close proximity and with a line of sight of the antenna. 
Otherwise, the adversary is forced to constantly track the position of the victim receiver and adapt its strategy to the trajectory of the victim, realistically simulating the propagation effects caused by the relative speed between transmitter and receiver. 
We consider an \textit{advanced} attacker that can lock adversarial signals to the GNSS signals, thus achieving a slow lift of the receiver from the legitimate GNSS without causing loss of tracking. The attacker can replay, replicate or alter data bits, spreading code and carrier information for any GNSS satellite; in other words, we do not restrict the attack techniques. Additionally, the adversary can produce code, phase and Doppler matched signals with gradually increasing power, obtaining control of the tracking loops using signal-lifting techniques. In practice, we assume that the attacker has complete control of the RF interface of the victim GNSS receiver, either with a physical connection or transmitting adversarial signals over the air. Finally, we assume that the attacker can be already active when the receiver is in cold start (initial acquisition of GNSS PVT). Due to the likely higher transmission power of the attacker, the victim receiver will likely lock to the adversarial signals. This is a strong model, aligned with the implementation in \cref{sec:implementation-experimental}. Showing effective detection against such a strong adversary implies effectiveness against weaker adversaries too.

We assume that the attacker is time-bounded in its interaction with the victim. Ideally, an attacker could cause arbitrarily small modifications to the victim's PVT solution to avoid triggering any attack detection mechanism. Practically, such a slow build-up approach might not be at all feasible if, for example, the victim receiver is mobile. Furthermore, attackers that have a prolonged interaction with the victim receiver might be detected by other means, such as RF spectrum monitoring or other (physical) intrusion detection methods. 

In this work, publicly available attack scenarios (TEXBAT, \cite{Humphreys2012a}) are used, which involve static or mobile receivers, with \textit{advanced} GNSS attackers, focusing on position or time tampering, capable of synchronized signal-lifting. Each attack scenario is also complemented by a ground truth case of absence of adversary, used as a validation case. 

\section{Proposed approach}
\label{sec:proposed-approach}
\begin{figure}[h]
\centering
  \includegraphics[width=0.7\linewidth]{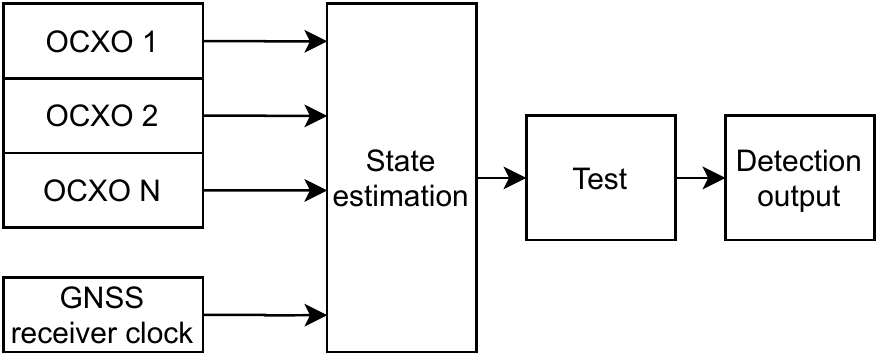}
  \caption{Multiple clocks can be used to test the GNSS receiver clock state and detect attacks tampering with the GNSS PVT.}
  \label{fig:system-model}
\end{figure}

The intuition of the attack detection approach elaborated on here is that GNSS attacks cause deviations that cannot be explained by the GNSS receiver's clock model. We specifically propose to test the offset and drift of the GNSS clock as part of an ensemble of high-stability oscillators, built in the enhanced receiver we propose in this work (Figure \ref{fig:system-model}).

Based on the clock model (Eq. \ref{eq:clock-model-diff}), we build a Kalman filter that tracks the GNSS receiver clock and the hardware oscillators. The model for tracking N clocks is defined by the following matrix-vector equation, at epoch $t$ and sampling step $\tau$:

\begin{equation}
    X(t) = \Phi(t)X(t-\tau) + W(t)
    \label{eqn:kf-1}
\end{equation}

where $X(t)$ is the vector of the clock states, x(t);  for N clocks this is a 3N vector: 
\begin{equation}
X(t)= [\theta_1(t), \gamma_1(t), D_1(t), ..., \theta_n(t), \gamma_n(t), D_n(t)]
\end{equation}

The transition matrix, $\Phi(\tau)$, is a 3N*3N matrix constructed with diagonal blocks $\phi(\tau)$:

\begin{equation}
\setstackgap{L}{1.1\baselineskip}
\fixTABwidth{T}
\phi(\tau) = \parenMatrixstack{
    1 & \tau & \frac{\tau^2}{2}\\
    0 & 1 & \tau\\
    0 & 0 & 1
}
\label{eqn:kf-2}
\end{equation}

The covariance matrix Q for the noise vector $W=[W_1(t), ..., W_n(t)]^T$ is 3N*3N diagonal matrix with blocks $q(t)$ defined in \cite{JonesTryonTimeSeries} as \cref{eqn:kf-3}, where the quantities $q_{\theta}, q_{\gamma}, q_{D}$ are the spectral noise densities of the process obtained from the Hadamard variance, according to \cite{Hutsell1995Relating}, for each clock.

\begin{equation}
\setstackgap{L}{1.1\baselineskip}
\fixTABwidth{F}
q(t) = \bracketMatrixstack{
    q_{\theta}\tau + \frac{q_{\gamma}\tau^3}{3} + \frac{q_{D}\tau^5}{20} & * & * \\
    \frac{q_{\gamma}\tau^2}{2} + \frac{q_{D}\tau^4}{8} & q_{\gamma}\delta + \frac{q_{D}\tau^3}{3} & * \\
    \frac{q_{D}\tau^3}{6} & \frac{q_{D}\tau^2}{2} & q_{D}\tau
}
\label{eqn:kf-3}
\end{equation}

The measurements equation for the filter is defined as:
\begin{equation}
    Z = HX(t)
    \label{eqn:kf-4}
\end{equation}

where the measurements matrix $H$ is build with N-1 rows of +1 and -1 for each clock the GNSS clock is compared to, and 0 elsewhere \cite{JonesTryonTimeSeries}. Hence, the noise-free measurements of the clock differences at epoch $t$ for $N$ clocks are expressed as:
\begin{equation}
\begin{tabular}{Lr}
x_{GNSSi}(t) = x_{GNSS}(t) - x_i(t) & with $i=1,..,N$
\end{tabular}
\label{eqn:kf-5}
\end{equation}

From \crefrange{eqn:kf-1}{eqn:kf-5} we build a discrete Kalman filter as indicated in \cite{JonesTryonTimeSeries}, where $t$ indicates the current epoch, and $t+1$ the next epoch (at sampling rate $\tau$):\\
\textbf{Prediction:}
\begin{equation}
    \begin{gathered}
        X(t+1|t) = \Phi(t+1)X(t|t) \\
        P(t+1|t) = \Phi(t+1)P(t|t)\Phi^T(t+1) + \tau Q(t+1) \\
    \end{gathered}
\end{equation}

\textbf{Innovation:}\\
\begin{equation}
    \begin{gathered}
        I(t+1) = Z(t+1) - Z(t+1|t) \\
        C(t+1) = H(t+1)P(t+1|t)H^T(t+1)+R(t+1) \\
    \end{gathered}
\end{equation}

\textbf{Update:}\\
\begin{equation}
    \begin{gathered}
        KG = P(t+1|t)H^T(t+1)C^{-1}(t+1) \\
        X(t+1|t+1) = X(t+1|t+1) + KG*I(t+1) \\
        P(t+1|t+1) = P(t+1|t) - KG*H(t+1)*P(t+1|t) \\
    \end{gathered}
\end{equation}

We initialize at time $t_0$ the state vector to $X_0 = [0, ..., 0]$ and $P_0$ to a identity matrix of dimensions 3N*3N. For local clocks we can assume measurement noise $R=0$, although picking a small diagonal $R$ matrix increases numerical stability \cite{JonesTryonTimeSeries}.

\subsection{Attack Detection}

\label{section:attack-detection}
The spoofing test we propose relies on the differential state estimates produced by the Kalman filter. We are interested in monitoring the phase, as this allows the receiver to quantify the amount of adversary-induced shift in the PVT solution. Additionally, we consider the frequency offset as this is an indicator of the aggressiveness of the attack, meaning the speed at which the attacker changes the frequency offset with respect to the nominal oscillator frequency.

An attack is detected when the state estimate produced by the Kalman filter is outside the confidence interval calculated based on the receiver clock model. At each epoch, the filter estimates are tested against a distribution of the clock offset and frequency offset whose standard deviation $\sigma_\theta, \sigma_\gamma$ are obtained by a one-time calibration of the detection system. This can be obtained from historical reference measurements in the absence of an adversary. We consider a true positive (true negative) when the system correctly identifies an ongoing attack (the absence of an adversary). A false positive occurs when the system detects adversarial signals while there is none. A false negative is the event of a missed detection of an adversary. The detection threshold is obtained as $6\sigma_\theta, 6\sigma_\gamma$ of the reference case, considering that $6\sigma$ is a conservative threshold maintaining the level of false positives to a minimum.


For static GNSS receivers that can achieve better time accuracy, because their position is known accurately, a lower threshold can be adopted. Mobile receivers often observe a degradation of the time solution, requiring a more relaxed threshold. If the mobility of the receiver is unknown, the worst case scenario can be adopted for the detection threshold. Practically, either threshold could be chosen depending on the specific application and it can be obtained from GNSS receiver information in a benign setting.

\begin{figure}[ht]
  \centering
  \includegraphics[width=0.6\linewidth]{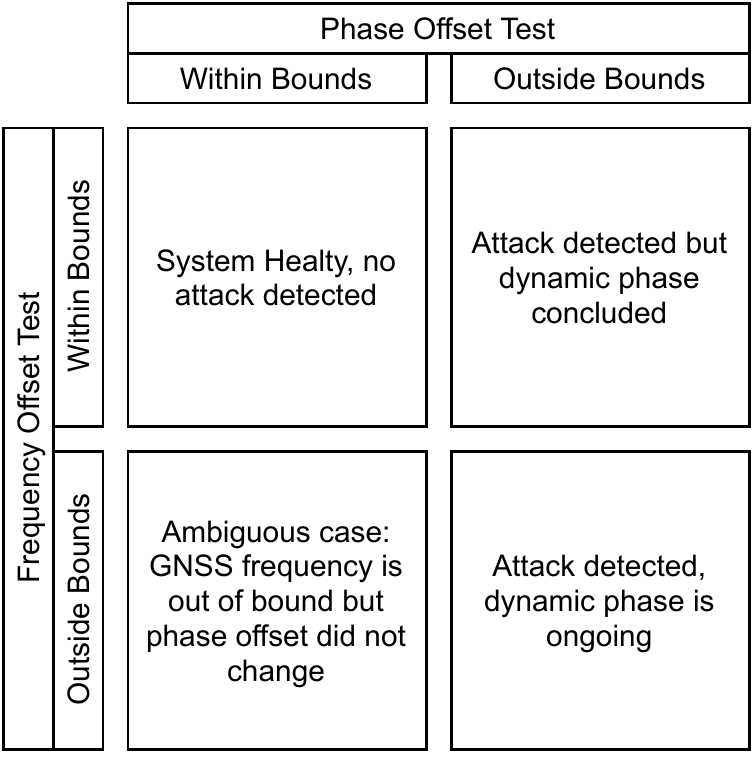}
  \caption{Possible decision outcomes based on time offset and frequency tests: values outside $6\sigma_\theta, 6\sigma_\gamma$ bounds indicate a possible attack.}
  \label{figure:test-matrix}
\end{figure}

Our method provides two tests towards detecting attacks on the GNSS receiver, one based on the phase offset, the other on the frequency offset of the GNSS receiver, estimated with the ensemble. The first test detects the onset of an attack and it can also be used to detect attacks that completed the dynamic phase, in which the victim's PVT solution is actively 'lifted' from the real one, while maintaining a constant victim time offset. 
The second test provides insights on potentially ongoing attacks, as it detects variations of the GNSS clock frequency that are beyond the stability model for a specific oscillator. Sudden variations in the receiver frequency provide insights on steering of the GNSS clock: too frequent or out-of-bound steering can be used to detect adversarial behaviour. This allows to not only detect attacks during the active lift phase, but to also determine the aggressiveness of the attacker action. To enhance the detection of an attacker, the two methods can be combined as shown in \cref{figure:test-matrix} where the possible outcomes of the tests are outlined. 

Intuitively, if both tests show that the estimates obtained from the ensemble for the GNSS receiver clock lie outside the $6\sigma$ confidence bounds, then the attack is considered detected and ongoing (the attacker is actively steering the victim's PVT solution). If only a time offset is measured in the PVT solution, the attack is partially detected and most likely caused by an adversary that reached its objective and maintains a stable offset for the victim receiver. If an attacker is present when the receiver is in cold start (and already reached its target time offset), the initial phase offset of the GNSS receiver clock with the real GNSS signals is unknown. This is a limitation of the current method, but a check with a trusted time provider (i.e, remote time server) addresses this problem. Nevertheless, the observation of the on-board clock ensemble remains a powerful tool to detect early attacks that are below the accuracy of the network time or in connection-denied environments.

Furthermore, if the frequency offset based test highlights an estimate out of the $6\sigma$ confidence bounds, but no time offset is detected the attacker detection might be ambiguous. Our countermeasure might cause a false positive in case the GNSS receiver steers its internal clock frequency to compensate for errors. On the other hand, abrupt changes in the internal clock frequency could still be of interest, towards detecting an attack, if they are outside the normal change rate provided by the manufacturer.

\section{Implementation and experimental setup}
\label{sec:implementation-experimental}

\label{section:clock-implementation}
To evaluate the attack detection capabilities of the proposed method, we implement a hardware prototype. The detection algorithm is designed and validated in MATLAB and runs on a processing core implemented in an FPGA alongside the hardware components required to precisely measure the phase and the frequency of the individual clocks. The algorithm is real-time and relies only on the current measurement plus the most recent estimate available to produce a decision.

\begin{figure}
    \centering
  \includegraphics[width=0.7\linewidth]{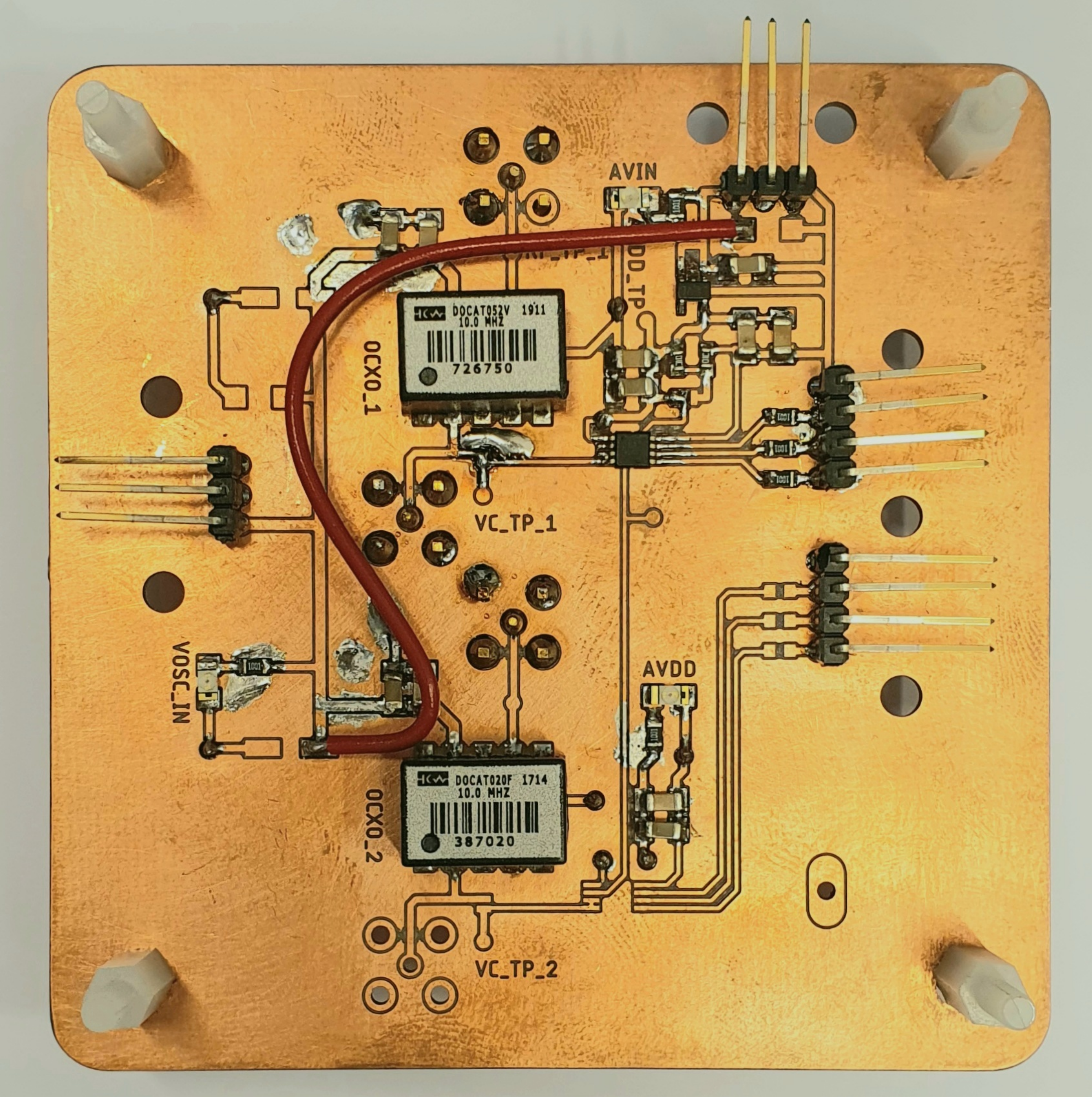}
  \caption{Clock carrier board: each clock has a decoupled power network, and individual clock buffer.}
  \label{fig:clock-carrier}
\end{figure}

Our system leverages up to four independent chip-scale oven-compensated oscillators with a typical Allan deviation of $\sigma_A(\tau)=$~\SI{5e-10} up to \SI{1e4}{\second}. A carrier board that provides power delivery and test connections is designed to host two clocks per each board, as shown in Figure \ref{fig:clock-carrier}. Decoupling of power networks and independent clock buffering allow keeping the clocks isolated.

The used GNSS receiver is the uBlox ZED-F9P dual frequency multi-constellation receiver. As the TEXBAT scenarios consider exclusively GPS signals, the system is configured to operate on the L1/L2 band with GPS satellites only. Although a more complex adversarial simulator could operate on multiple constellations at the same time, it does not limit the validity of the results presented, as the detection system is agnostic of the specific constellation (or their combination) used by the GNSS receiver to calculate the PVT solution. 

\begin{figure}[h]
    \centering
  \includegraphics[width=0.7\linewidth]{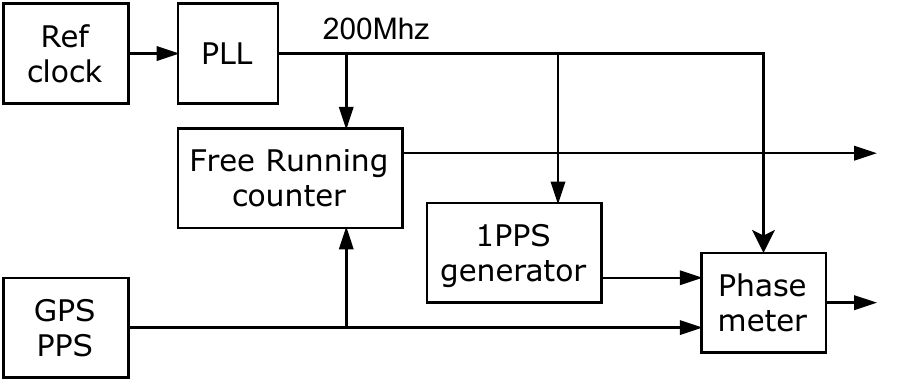}
  \caption{Basic reference clock block: the PLL is used to create high speed clocks that measure the phase of the GPS signal and provide precise timekeeping.}
  \label{fig:clock-block}
\end{figure}

The time-sensitive measurements of the attack detection system are implemented on an FPGA (Figure \ref{fig:clock-block}). This solution combines time-sensitive phase measurement logic with embedded micro-controllers in a small form-factor, allowing reconfiguration and scaling of the system, potentially growing the ensemble, with more clocks. The phase offset between the 1-pulse per second (1PPS) line obtained from the GNSS receiver and a 1PPS derived from each reference clock is measured by means of an asynchronous phase detector \cite{razavi2000}. This allows measuring the phase shift with a resolution limited only by the counter clock. The phase information is processed by tracking the time and frequency information obtained by the GNSS receiver.

\begin{figure}[h]
    \centering
  \includegraphics[width=0.5\linewidth]{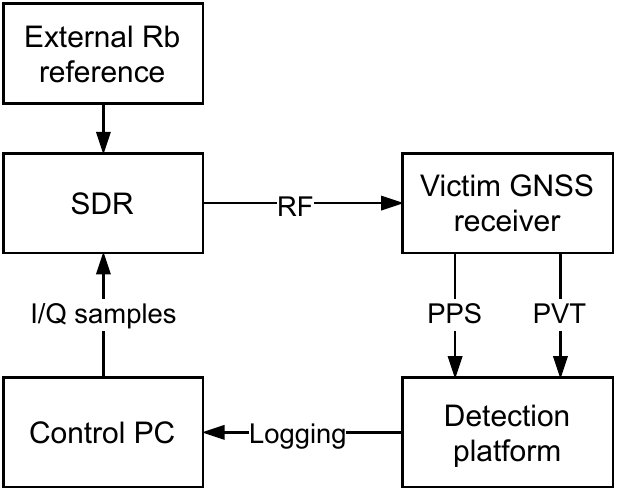}
  \caption{Setup used to evaluate different adversarial scenarios. The SDR timing is provided by a precision rubidium (Rb) oscillator.}
  \label{fig:adversary-player}
\end{figure}

\begin{figure}[h]
    \centering
  \includegraphics[width=0.8\linewidth]{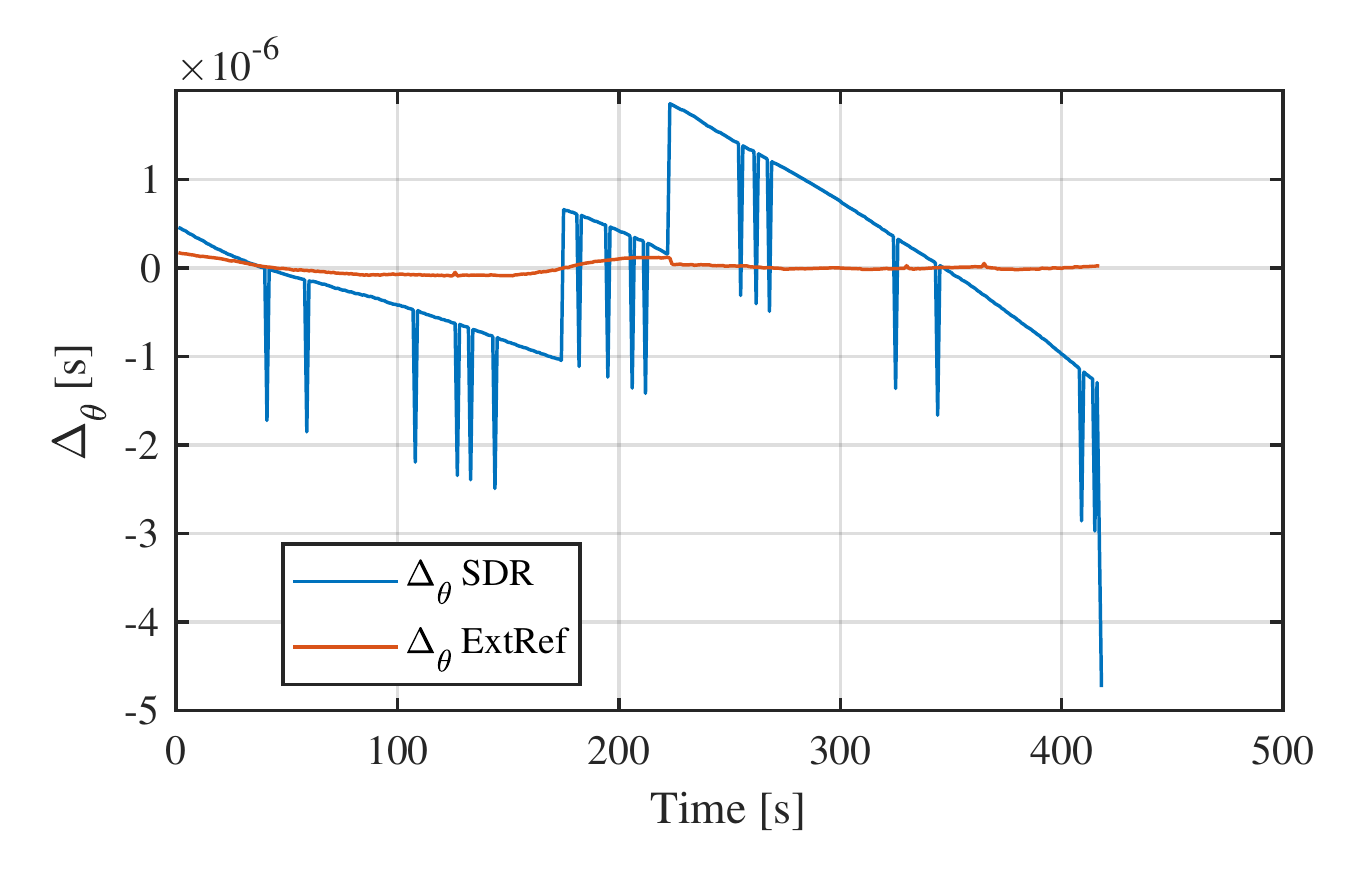}
  \caption{Phase error, $\Delta_\theta$, due to the SDR on-board clock. The error is removed when a rubidium oscillator is used as external reference.}
  \label{fig:phase-error-extref}
\end{figure}

To generate the spoofing scenarios, a BladeRF software-defined radio (SDR) is used to 'play' the scenarios from TEXBAT \cite{Humphreys2012}, allowing repeatable evaluation of various GPS-focused spoofing attacks, as shown in Figure \ref{fig:adversary-player}. 
Although the SDR clock is suitable to generate GNSS signals that allow the receiver to calculate the correct time and position, the quality of the radio's clock is critical to achieve good stability of the generated PPS information. This is evident in the phase plot in Figure \ref{fig:phase-error-extref}. The phase of the PPS signal is measured against a rubidium standard reference clock. First, when the SDR is clocked using the on-board oscillator, the phase presents a significant variance. Second, a rubidium oscillator is used as a reference for the SDR, resulting in variance one order of magnitude lower. In both cases, the phase is measured against the same reference, to remove potential errors due to the phase measurement system.

Due to imperfections of the Phase Locked Loop (PLL) used to generate the sampling clock and the local 1PPS reference for the phase stability measurement, a one-time calibration of the counter that generates the reference signal is required. Such calibration, achieved by using a known 1PPS reference, is device-dependant. We observed that without the calibration step, the local reference is subject to frequency beating that causes the phase measurement to drift linearly. The calibration is limited by the finite resolution of the counter, but further refinement can be achieved in software, by removing the fractional parts of the error.

\section{Evaluation}
\label{sec:evaluation}

We test our detection scheme against static and mobile adversarial scenarios obtained from the TEXBAT simulation files \cite{Humphreys2012}. Scenarios 2, 3 and 5 focus on the modification of time at the victim receiver. Specifically, the attacker operates aligning the spreading code and navigation information with legitimate signals, without locking the phase of the spoofing signal to the legitimate signal phase (Scenarios 2 and 5). From a technical standpoint, code phase alignment lock is also possible, but the spoofer complexity increases. An example of this is provided in Scenario 3, where a phased locked adversary overtakes a static GNSS receiver. In each experiment, the receiver obtained a clean GPS PVT before the attack. Each attack scenario starts at T=\SI{60}{\second}, with slight variations depending on the receiver's time to first fix (TTFF).

\begin{figure*}
  \centering
  \begin{subfigure}[t]{0.32\textwidth}
    \includegraphics[width=\linewidth]{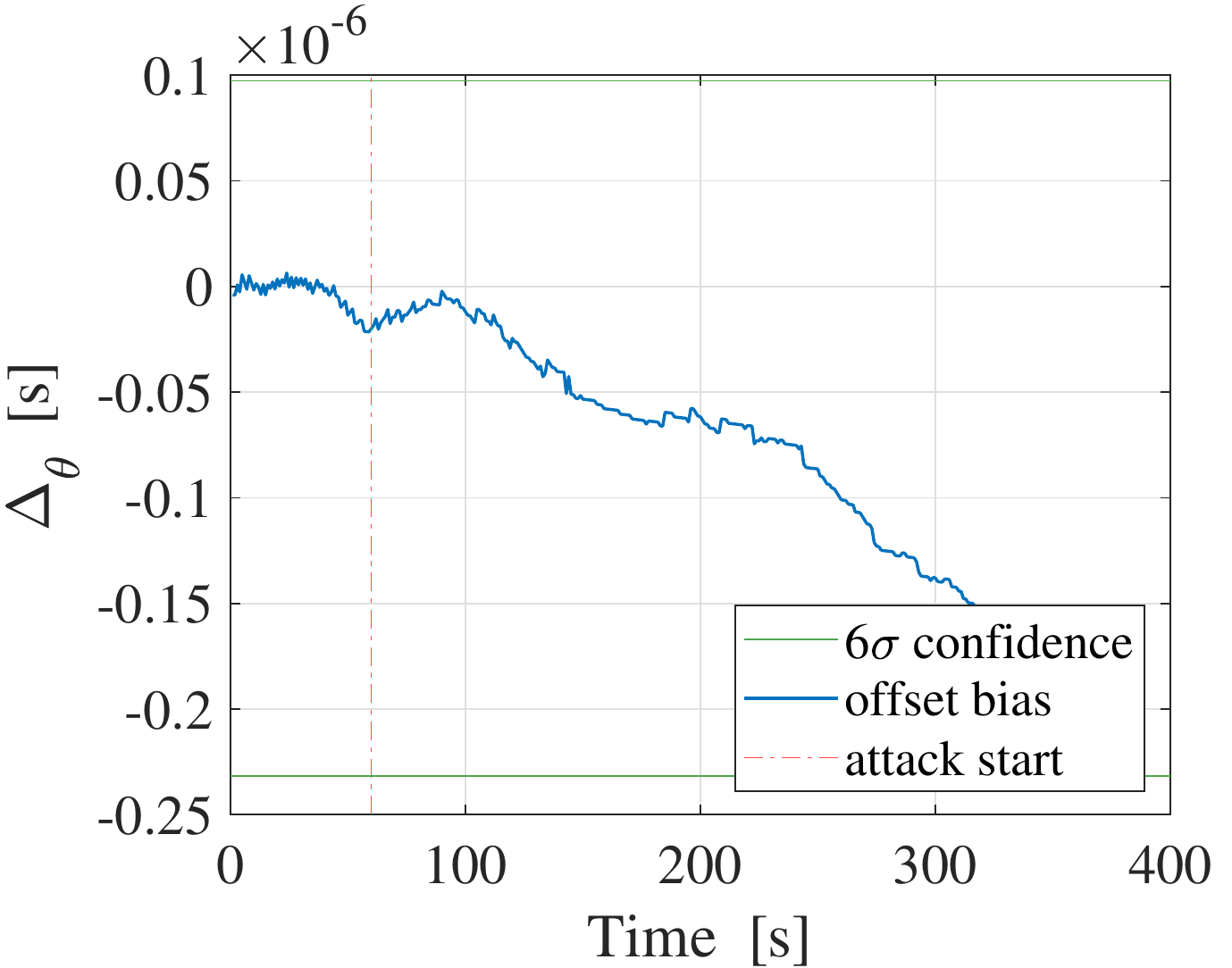}
    \caption{Static receiver without attack: the phase difference between GNSS clock and the ensemble is within tolerance.}
    \label{fig:time-offset-difference-ref-gps-clean-static}
  \end{subfigure}
  \hfill%
  \begin{subfigure}[t]{0.33\textwidth}
    \includegraphics[width=\linewidth]{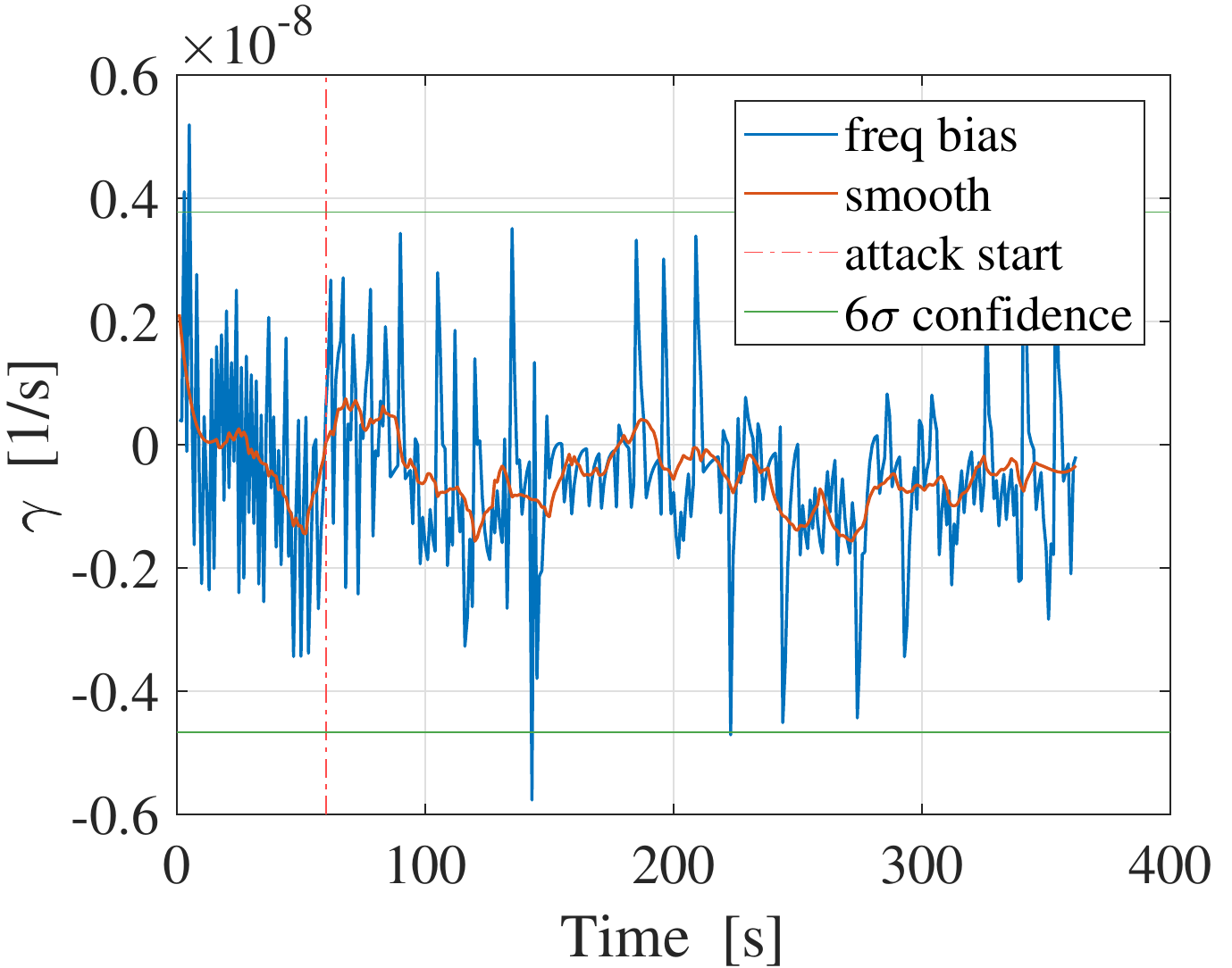}
    \caption{Frequency difference measured without attack: the frequency stability of the GNSS disciplined clock is within tolerance.}
    \label{fig:freq-offset-difference-ref-gps-clean-static}
  \end{subfigure}
  \hfill%
  \begin{subfigure}[t]{0.33\textwidth}
    \includegraphics[width=\linewidth]{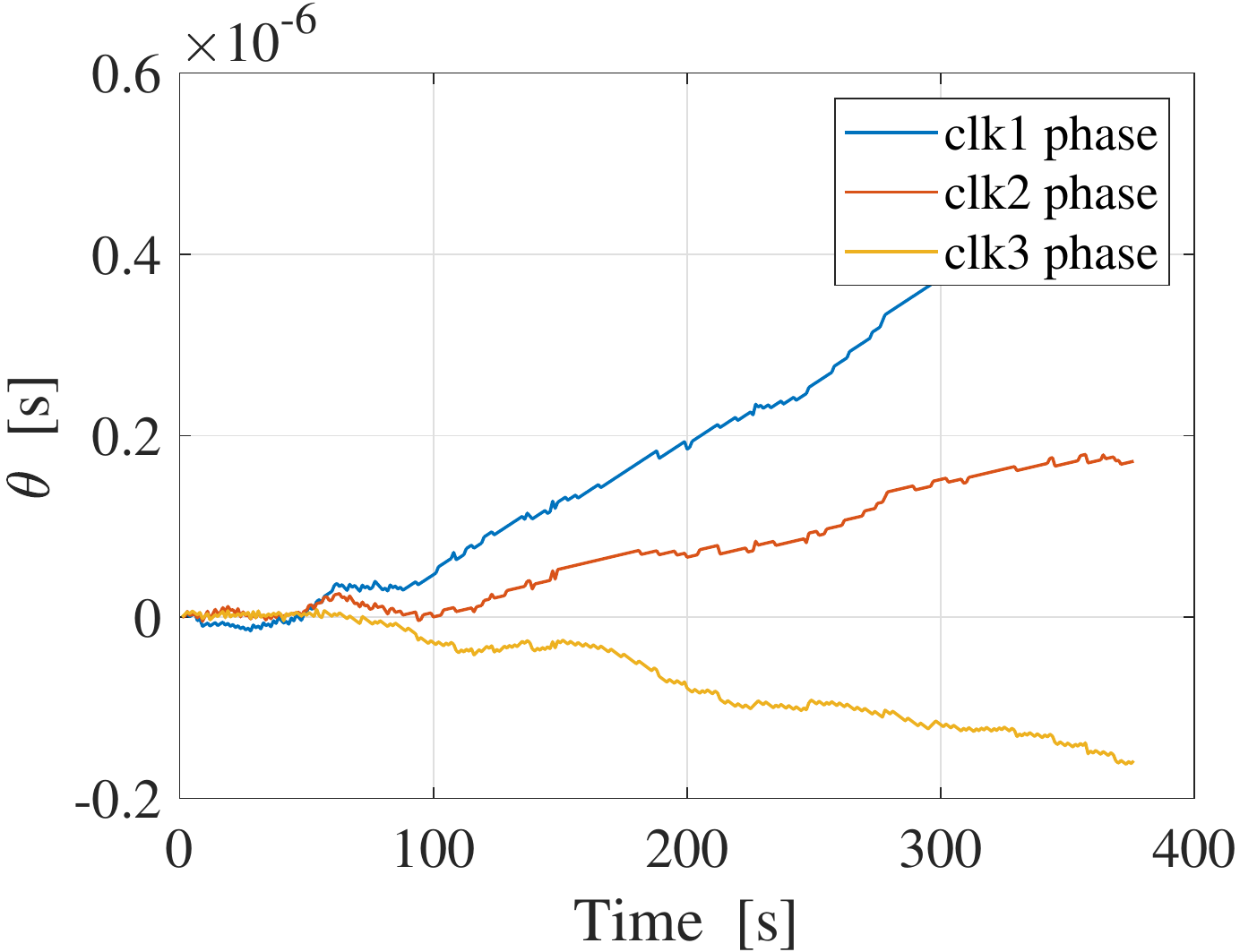}
    \caption{Phase difference corrections measured with three independent oscillators, without the ensemble.}
    \label{fig:zero-mean-phases-clean-static}
  \end{subfigure}
  \caption{Static reference scenario: static receiver without attack.}
   \label{fig:clean-static}
\end{figure*}

\begin{figure*}
  \centering
  \begin{subfigure}[t]{0.32\textwidth}
    \includegraphics[width=\linewidth]{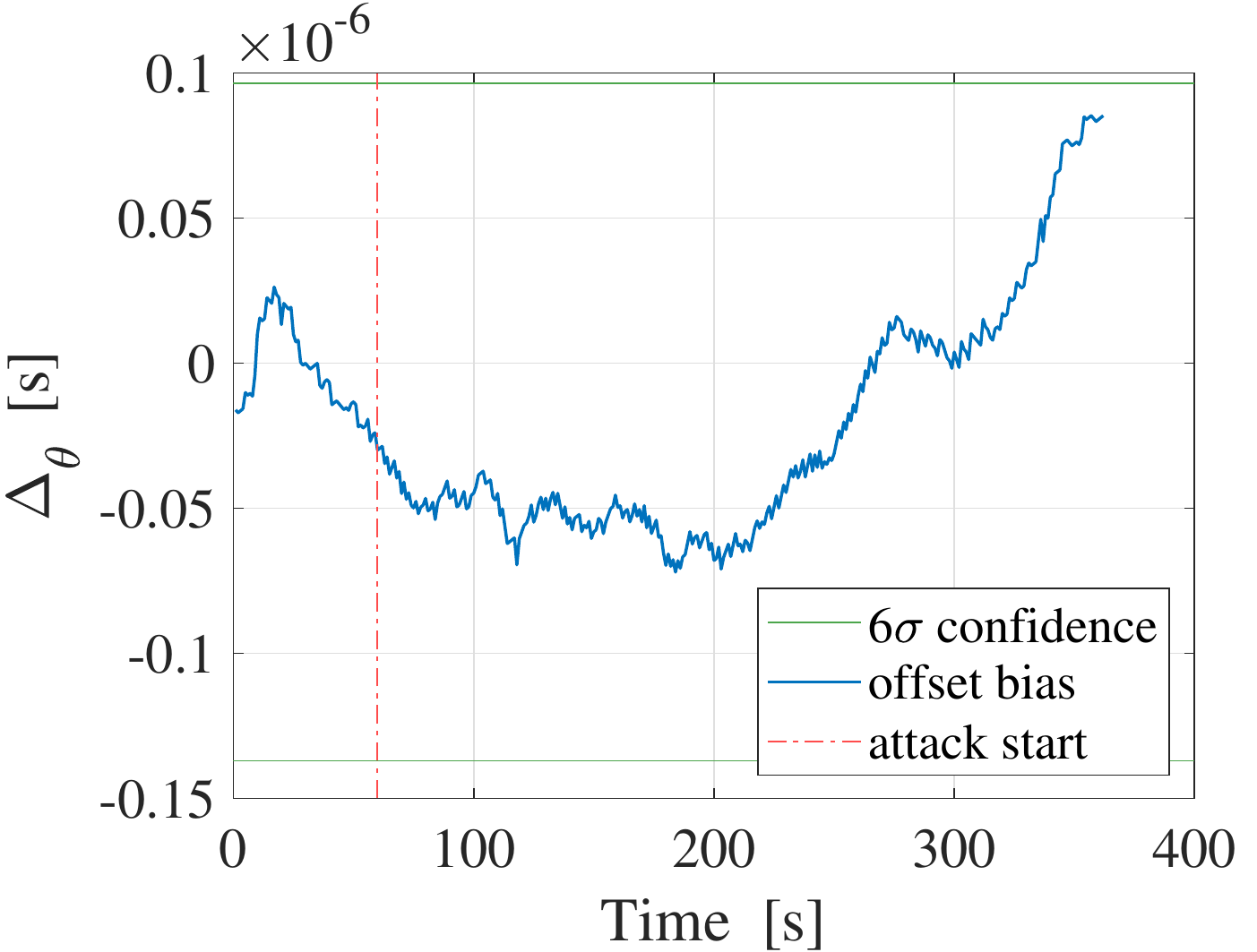}
    \caption{Mobile scenario without attack: the phase difference between GNSS clock and the ensemble is within tolerance.}
    \label{fig:time-offset-difference-ref-gps-clean-dynamic}
  \end{subfigure}
  \hfill%
  \begin{subfigure}[t]{0.33\textwidth}
    \includegraphics[width=\linewidth]{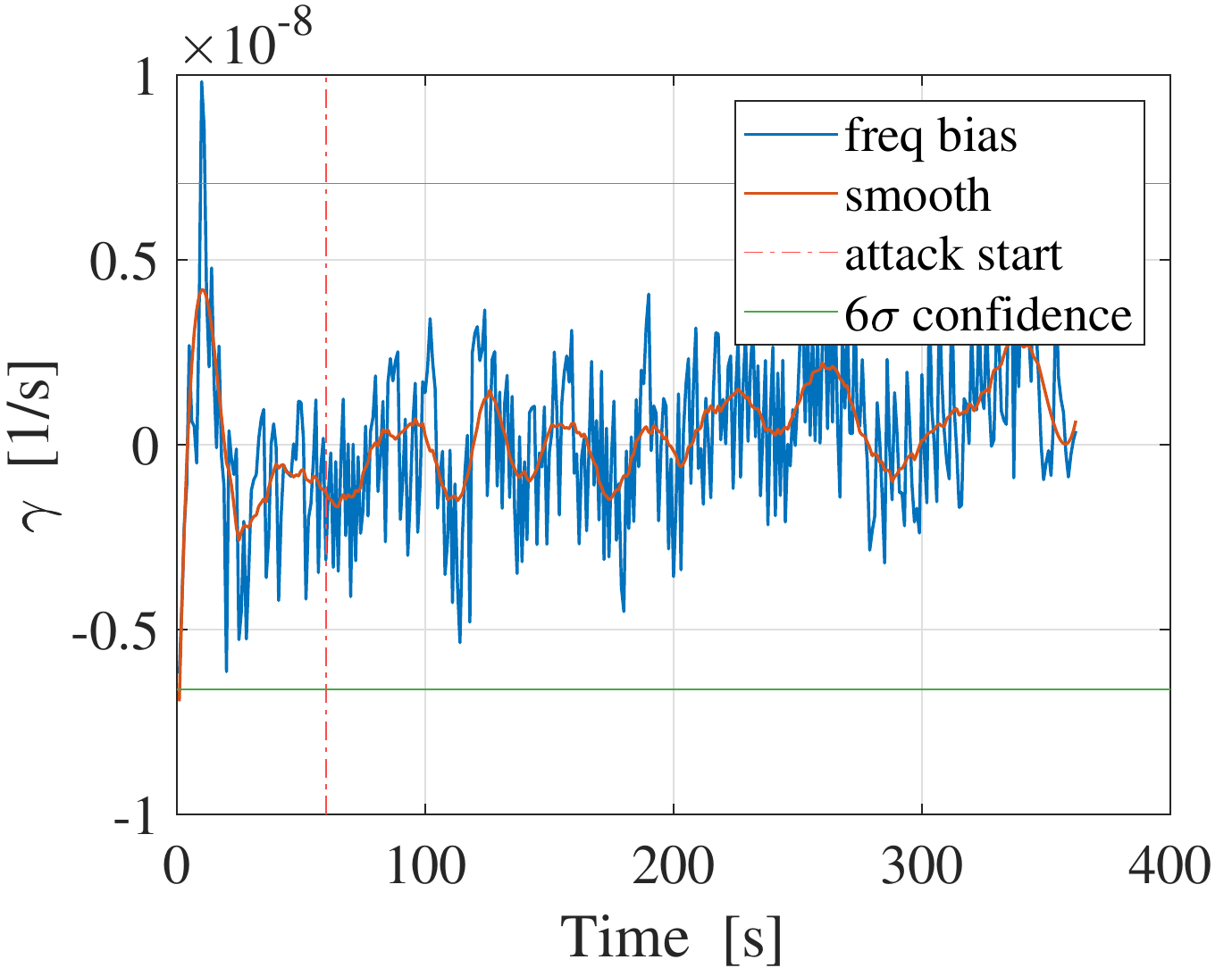}
    \caption{Frequency difference measured without an attack: the frequency stability of the GNSS disciplined clock is within tolerance.}
    \label{fig:freq-offset-difference-ref-gps-clean-dynamic}
  \end{subfigure}
  \hfill%
  \begin{subfigure}[t]{0.33\textwidth}
    \includegraphics[width=\linewidth]{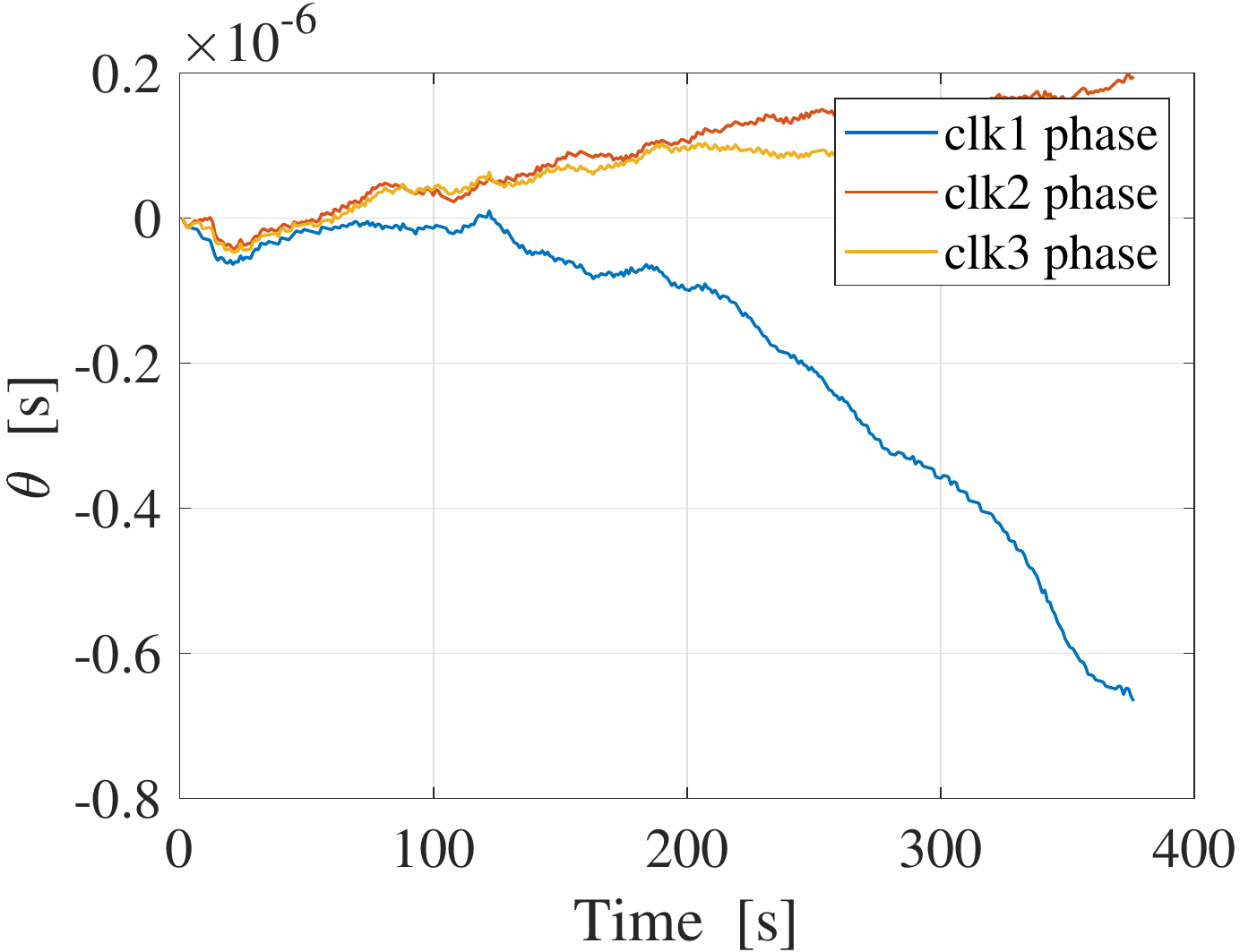}
    \caption{Phase difference corrections measured with three independent oscillators, without the ensemble.}
    \label{fig:zero-mean-phases-clean-dynamic}
  \end{subfigure}
  \caption{Mobile reference scenario: static receiver without attack.}
   \label{fig:clean-dynamic}
\end{figure*}

Figure \ref{fig:clean-static} shows the behaviour of the system for a static receiver in the absence of an attacker. The phase offset standard deviation, $\sigma_\theta=$~\SI{5.5834e-08}, is used as a base for the detection threshold of the phase offset test and the frequency offset standard deviation, $\sigma_\gamma=$~\SI{1.4109e-09}, is used as a base for the detection threshold of the frequency offset test. \cref{fig:clean-dynamic} shows the behaviour of the system for a mobile receiver in the absence of an attacker, where $\sigma_\theta=$~\SI{3.5606e-08} and $\sigma_\gamma=$~\SI{2.2561e-09}. As mentioned in \cref{section:attack-detection}, a threshold of $6\sigma$ is chosen, to minimize the false positive rate. 
 
The metrics we use to evaluate our system are total offset at the detection threshold and latency of detection. The former defines the amount of adversary-induced time offset at the victim when an attack is detected, the latter is defined as the amount of time that passes between the start of the attack and the detection system rising an alarm. 
Lower thresholds, e.g., $4\sigma$ would reduce the detection latency as long as the application is able to tolerate a potentially higher number of false positives.


\begin{figure*}
  \centering
  \begin{subfigure}[t]{0.32\textwidth}
    \includegraphics[width=\linewidth]{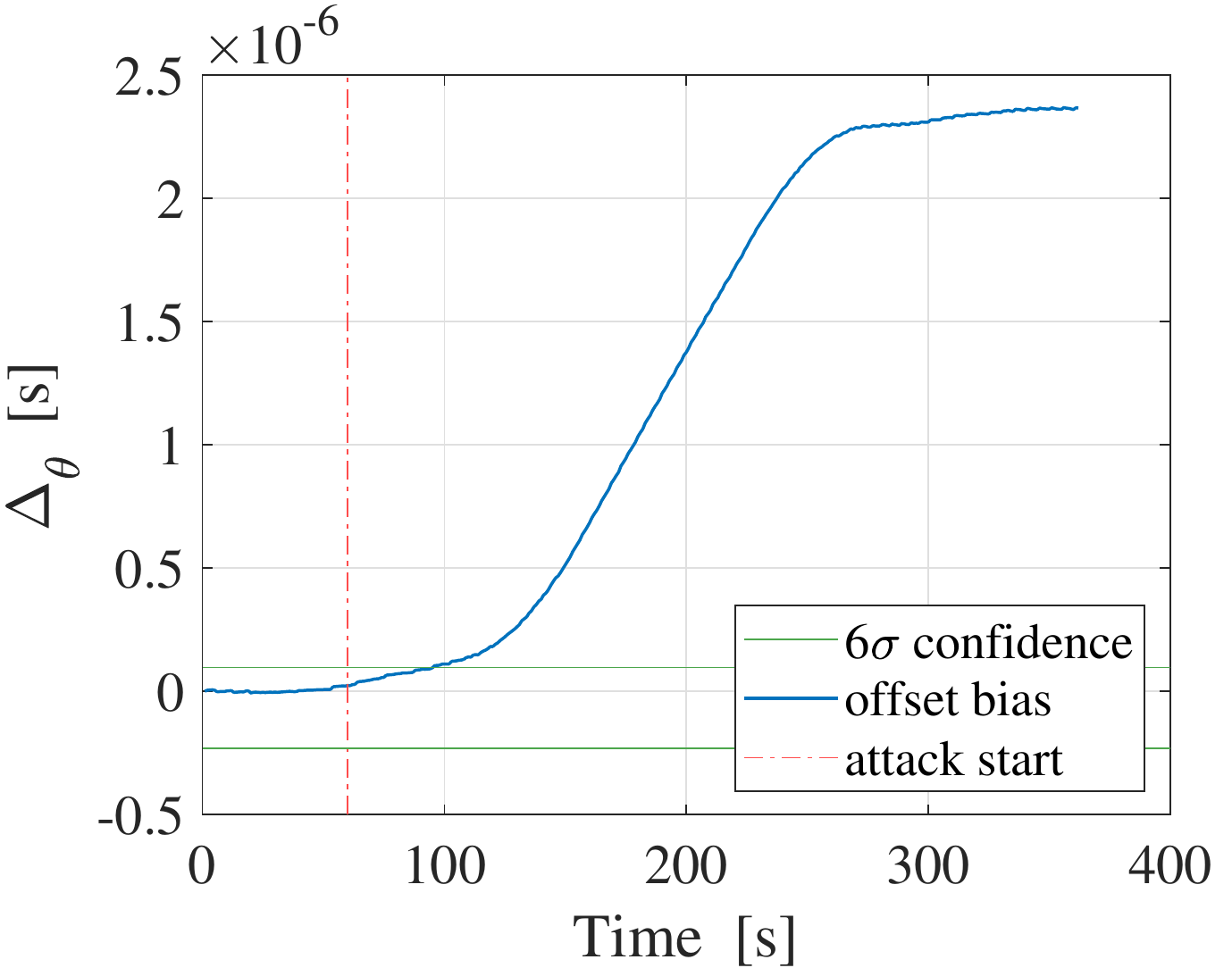}
    \caption{Phase difference measured when under attack: attacker induced \SI{2}{\micro\second} error starting at T=\SI{60}{\second}.}
    \label{fig:time-offset-difference-ref-gps-ds-2}
  \end{subfigure}
  \hfill%
  \begin{subfigure}[t]{0.32\textwidth}
    \includegraphics[width=\linewidth]{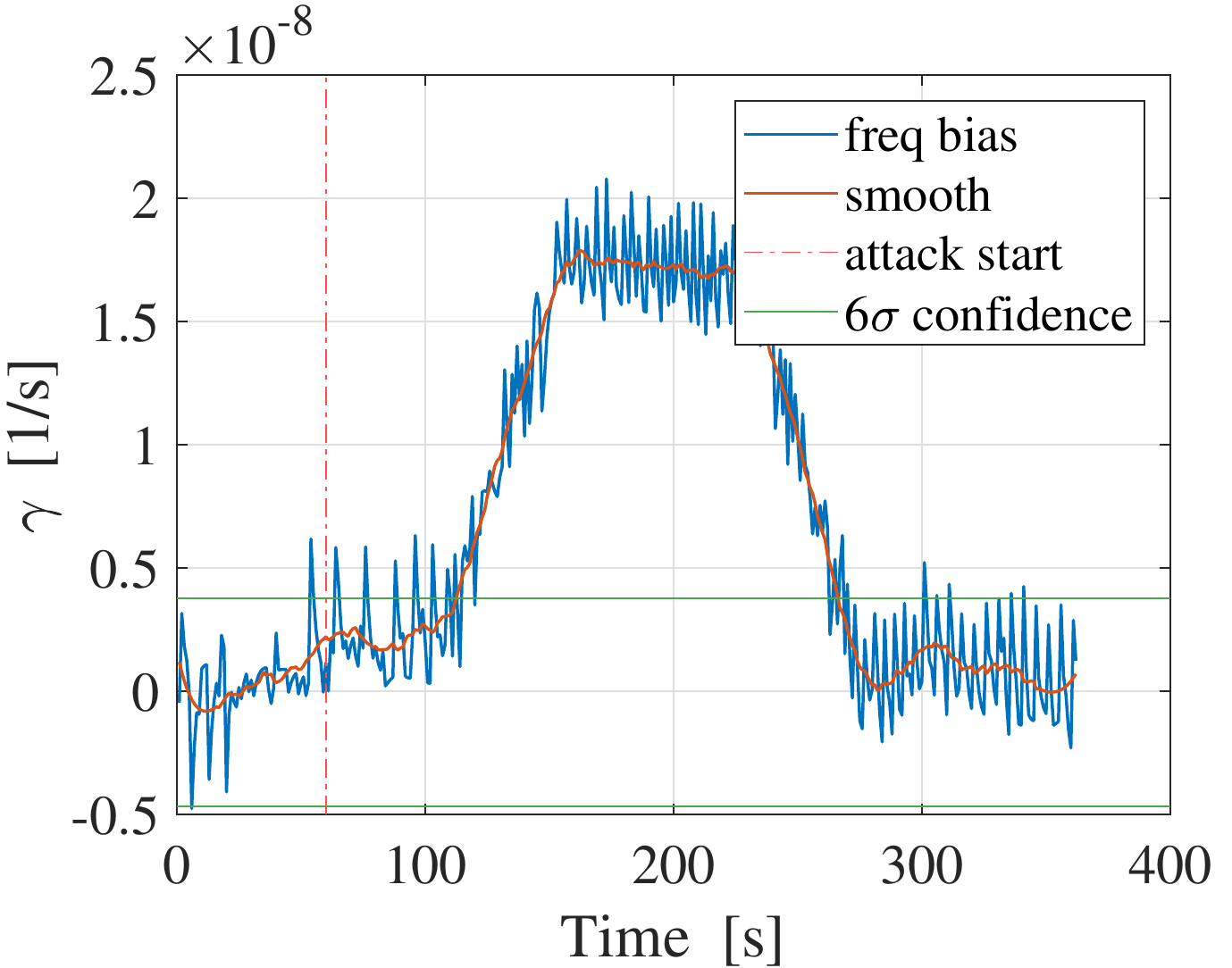}
    \caption{Frequency difference measured when under attack: GNSS-disciplined clock frequency subject to a frequency pull, up to \SI{20}{\nano\second/\second}.}
    \label{fig:freq-offset-difference-ref-gps-ds-2}
  \end{subfigure}
  \hfill%
  \begin{subfigure}[t]{0.33\textwidth}
    \includegraphics[width=\linewidth]{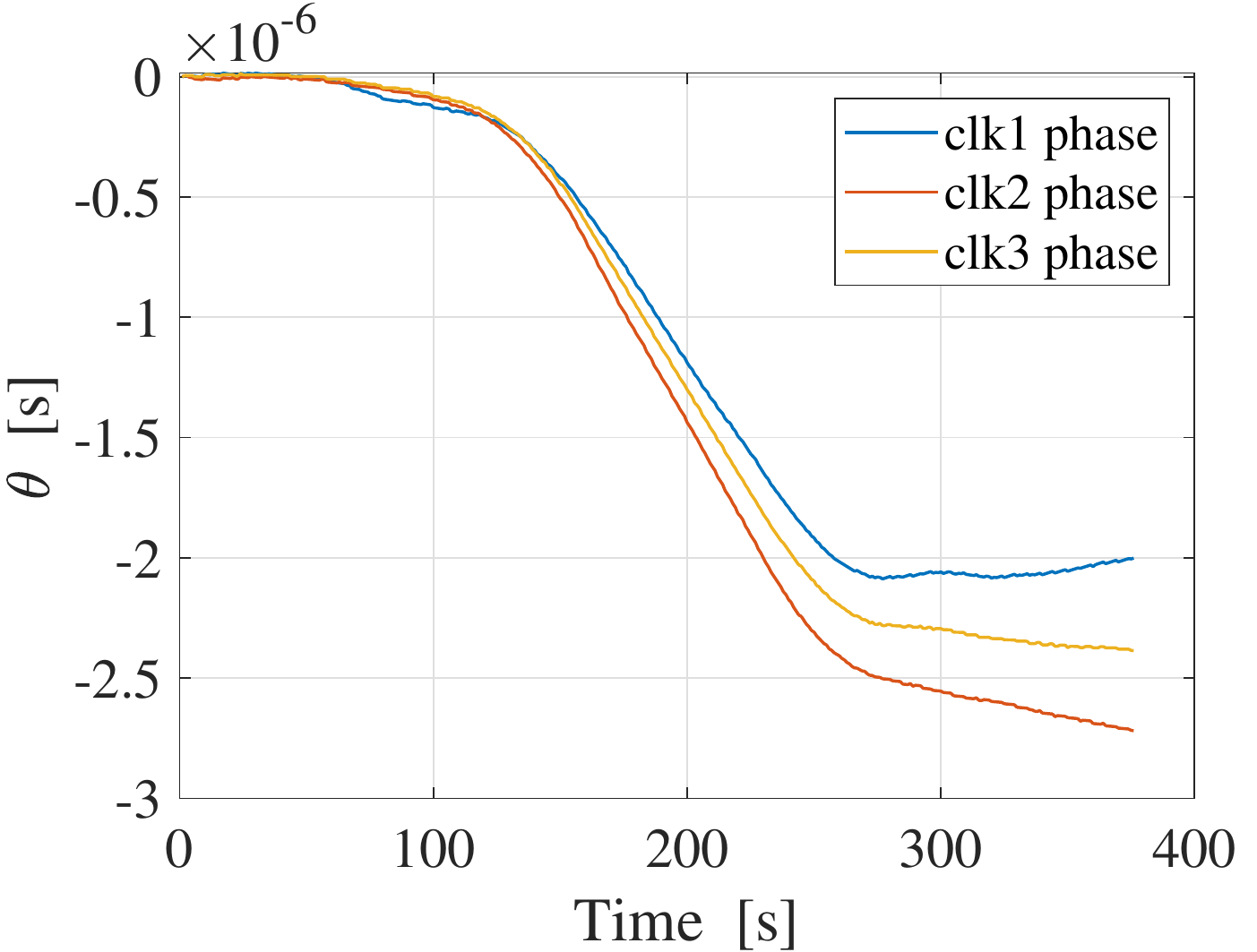}
    \caption{Phase difference correction measured with three independent oscillators. Each clock detects the attack independently.}
    \label{fig:zero-mean-phases-ds-2}
  \end{subfigure}
  \caption{Static receiver and phased unlocked adversary (Scenario 2).}
  \label{fig:spoofed-ds-2}
\end{figure*}

\begin{figure*}
\begin{subfigure}[t]{0.45\textwidth}
\centering
    \includegraphics[width=0.65\linewidth]{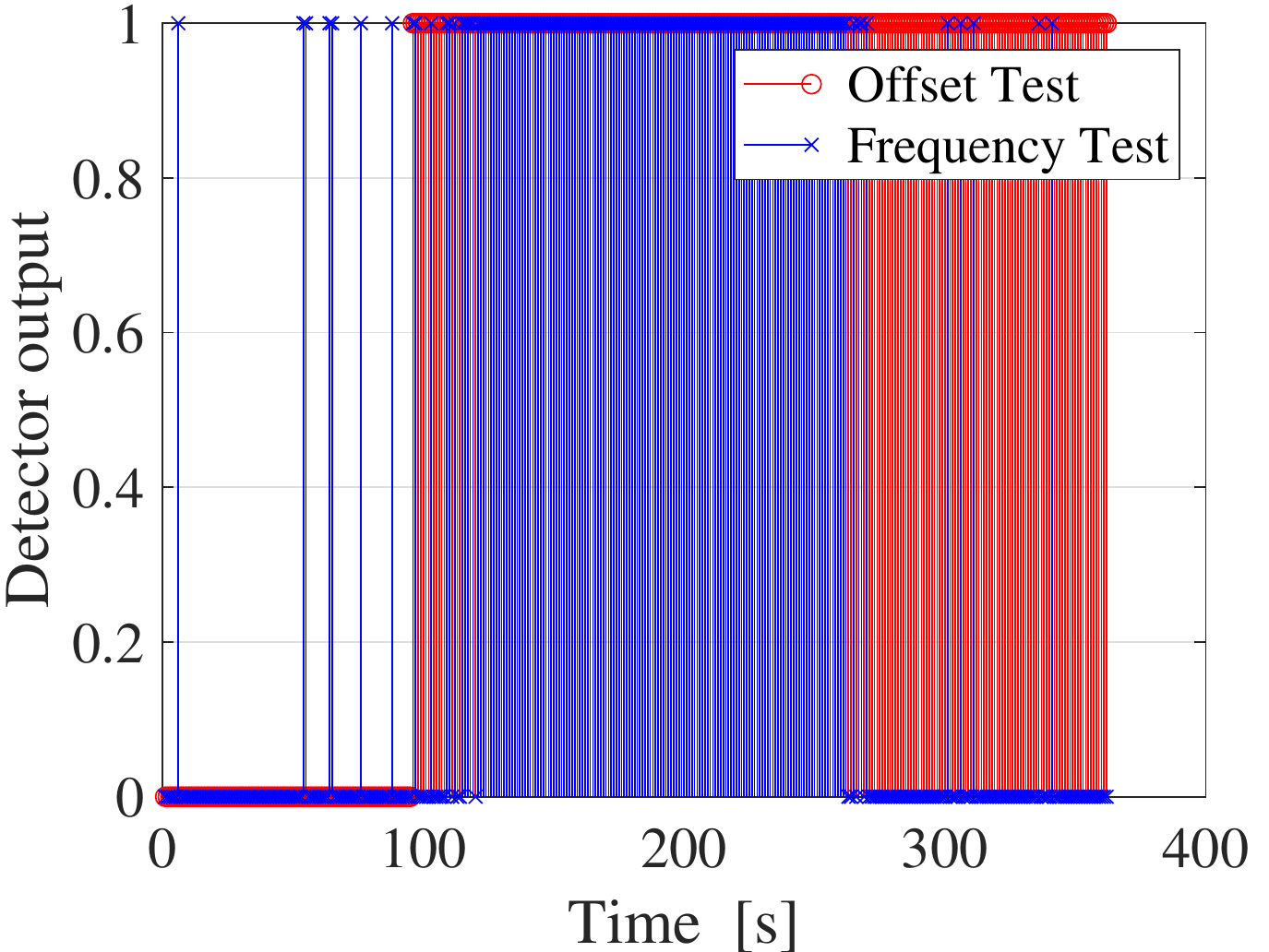}
    \caption{Static spoofed receiver: phase and frequency test results.}
    \label{fig:both-detector-output-ds-2}
  \end{subfigure}
  \hfill%
\begin{subfigure}[t]{0.45\textwidth}
\centering
    \includegraphics[width=0.65\linewidth]{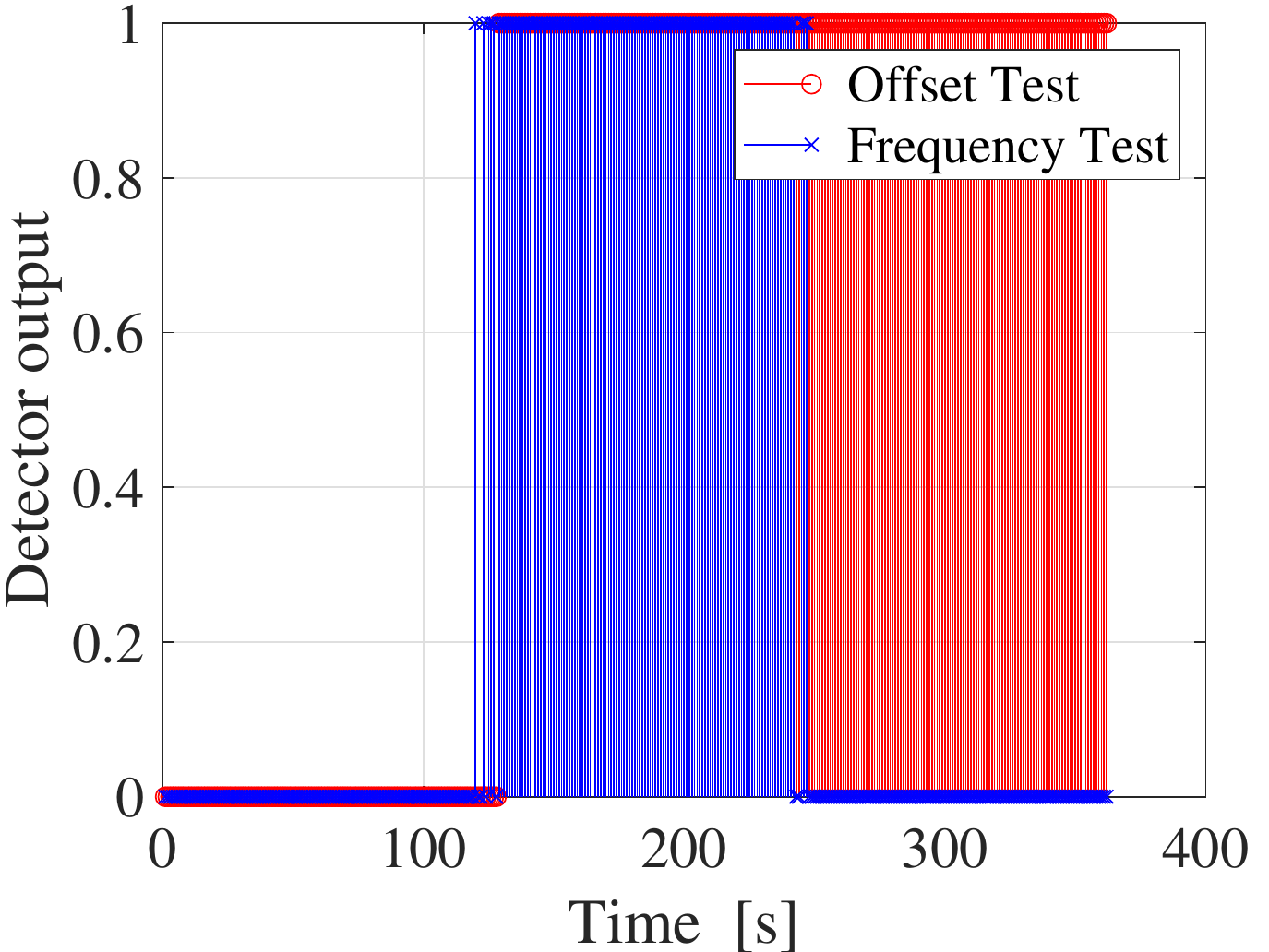}
    \caption{Dynamic spoofed receiver: phase and frequency test results.}
    \label{fig:both-detector-output-ds-5}
  \end{subfigure}
  \caption{Detector output for spoofed scenarios.}
  \label{fig:detector-output}
\end{figure*}

For a static receiver (Scenario 2), the phase offset estimate indicates a total of \SI{2}{\micro \second} change induced by the attacker (Figure \ref{fig:time-offset-difference-ref-gps-ds-2}). The phase offset clearly deviates beyond the confidence interval defined by $6\sigma_\theta$ and stabilizes around the new offset defined by the attacker, triggering the phase offset test at T=\SI{100}{\second}. Additionally, the frequency pull shown in Figure \ref{fig:freq-offset-difference-ref-gps-ds-2} allows the detection system to measure in real time the rate at which the attacker is dragging the time solution.
Figure \ref{fig:both-detector-output-ds-2} shows the benefit of using a double test on the GNSS receiver clock state. Once the lift phase of the attack is concluded (this information is obtained from the frequency offset test) at T=\SI{250}{\second}, the offset test can be used to detect the persistence of the attacker.

\begin{figure*}
  \centering
  \begin{subfigure}[t]{0.32\textwidth}
    \includegraphics[width=\linewidth]{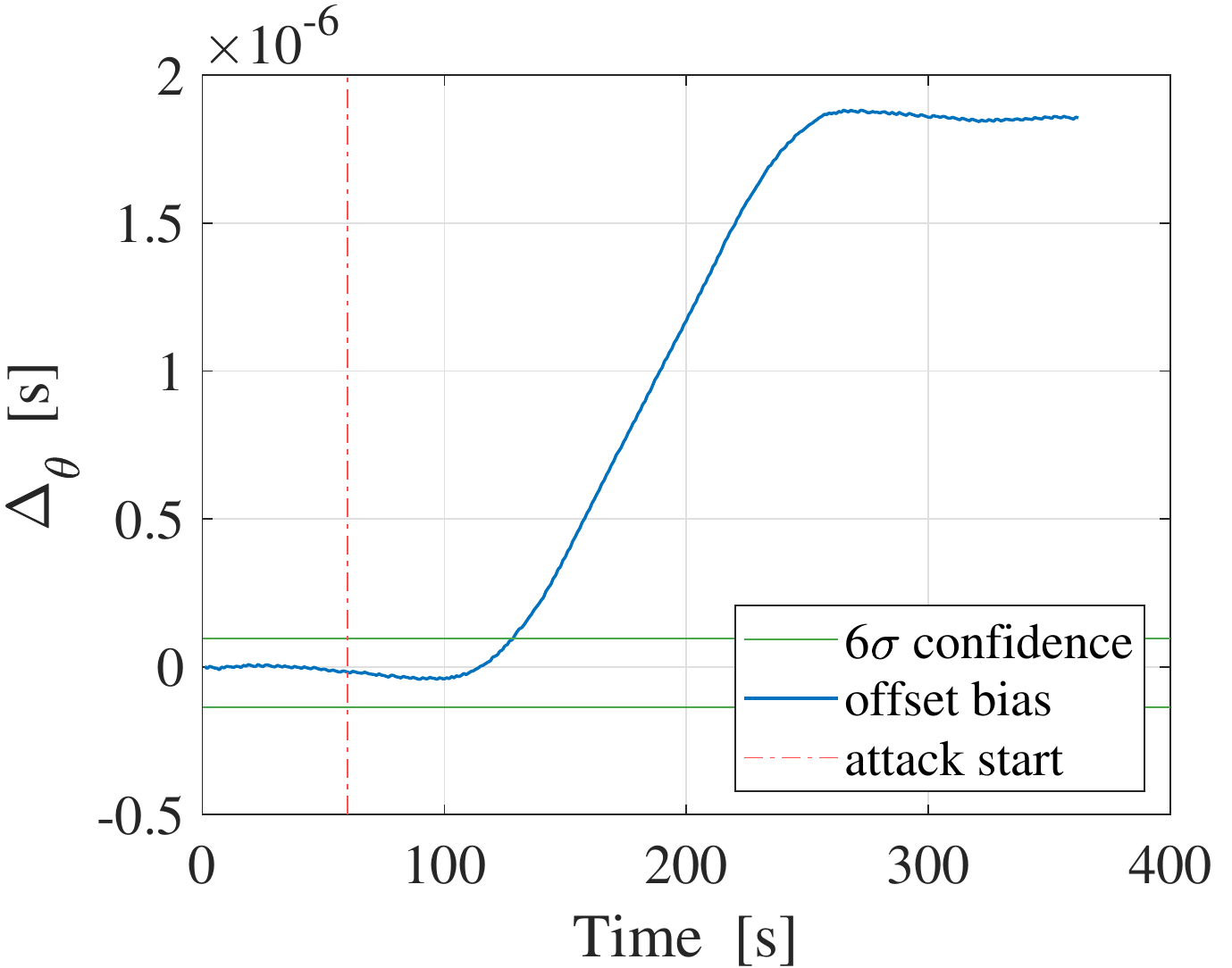}
    \caption{Phase difference measured when under attack: attacker induced \SI{2}{\micro\second} error starting at T=\SI{60}{\second}.}
    \label{fig:time-offset-difference-ref-gps-ds-5}
  \end{subfigure}
  \hfill%
  \begin{subfigure}[t]{0.32\textwidth}
    \includegraphics[width=\linewidth]{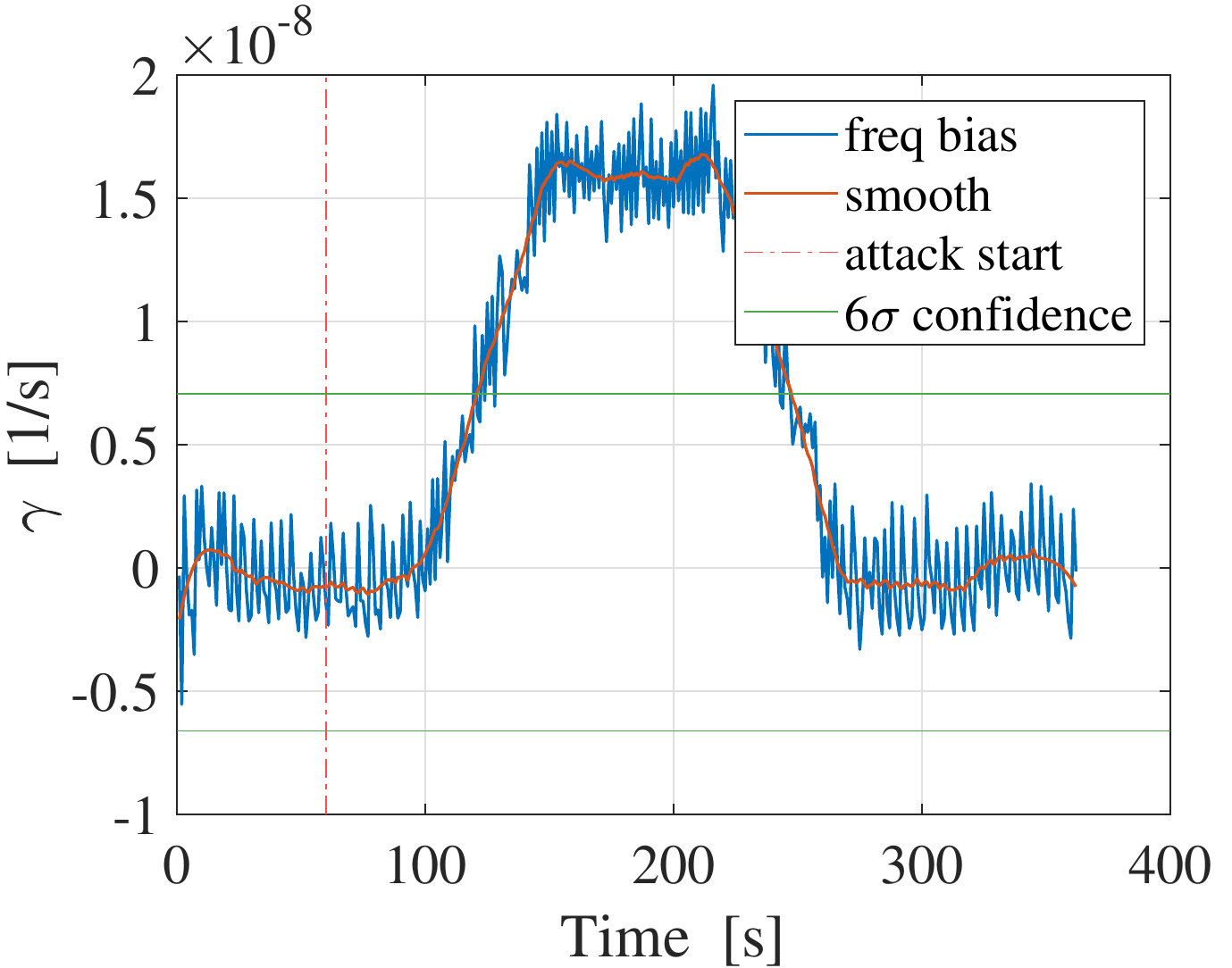}
    \caption{Frequency difference measured when under attack: GNSS-disciplined clock frequency subject to a frequency pull, up to \SI{20}{\nano\second/\second}.}
    \label{fig:freq-offset-difference-ref-gps-ds-5}
  \end{subfigure}
  \hfill%
  \begin{subfigure}[t]{0.33\textwidth}
    \includegraphics[width=\linewidth]{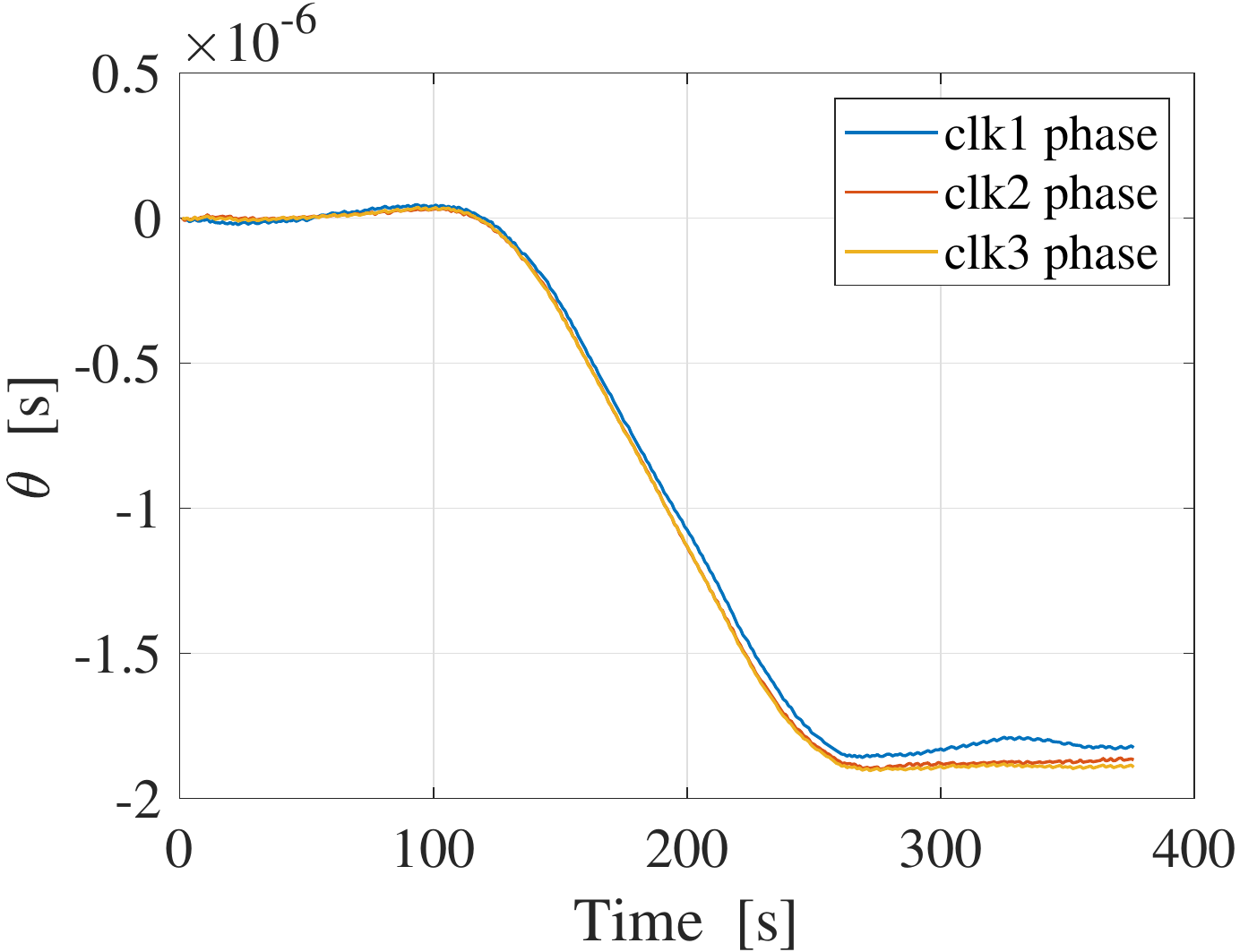}
    \caption{Phase difference correction measured with three independent oscillators. Each clock detects the attack independently.}
    \label{fig:zero-mean-phases-ds-5}
  \end{subfigure}
  \caption{Mobile receiver and phased unlocked adversary (Scenario 5).}
  \label{fig:spoofed-ds-5}
\end{figure*}

Similarly, for a mobile receiver (Scenario 5) the attack causes a total \SI{1.8}{\micro \second} phase offset, clearly visible in Figure \ref{fig:time-offset-difference-ref-gps-ds-5}. The frequency offset estimation shows that the lift-off phase is performed by the attacker between T=\SI{100}{\second} and T=\SI{250}{\second} (Figure \ref{fig:freq-offset-difference-ref-gps-ds-5}). Figure \ref{fig:both-detector-output-ds-5} shows the detector output for a spoofed dynamic receiver.

\begin{figure*}
  \centering
  \begin{subfigure}[t]{0.32\textwidth}
    \includegraphics[width=\linewidth]{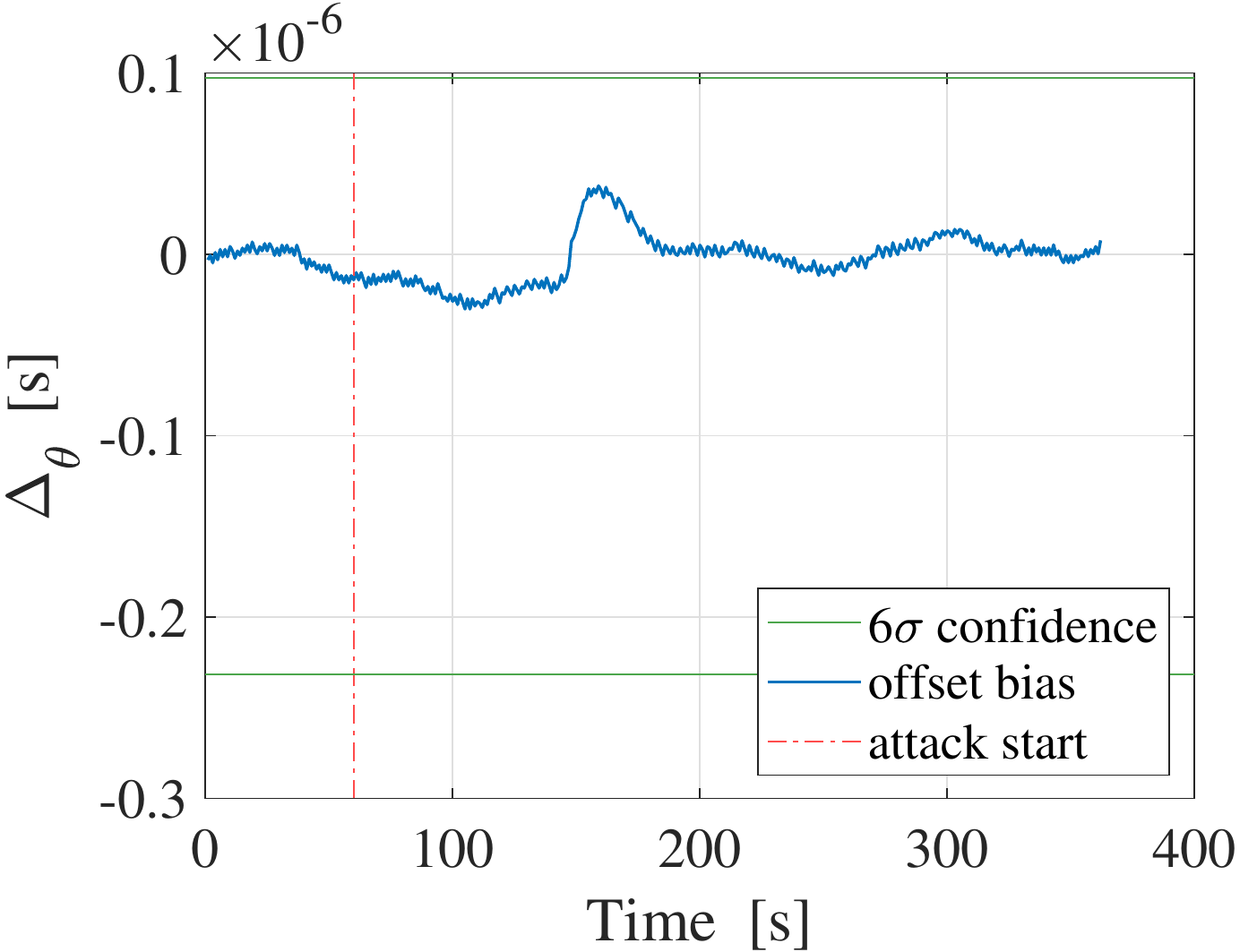}
    \caption{Phase difference measured when under attack and spoofer in phase locked mode. Time offset within the tolerance range, effectively not affecting time at the receiver.}
    \label{fig:time-offset-difference-ref-gps-ds-3}
  \end{subfigure}
  \hfill%
  \begin{subfigure}[t]{0.32\textwidth}
    \includegraphics[width=\linewidth]{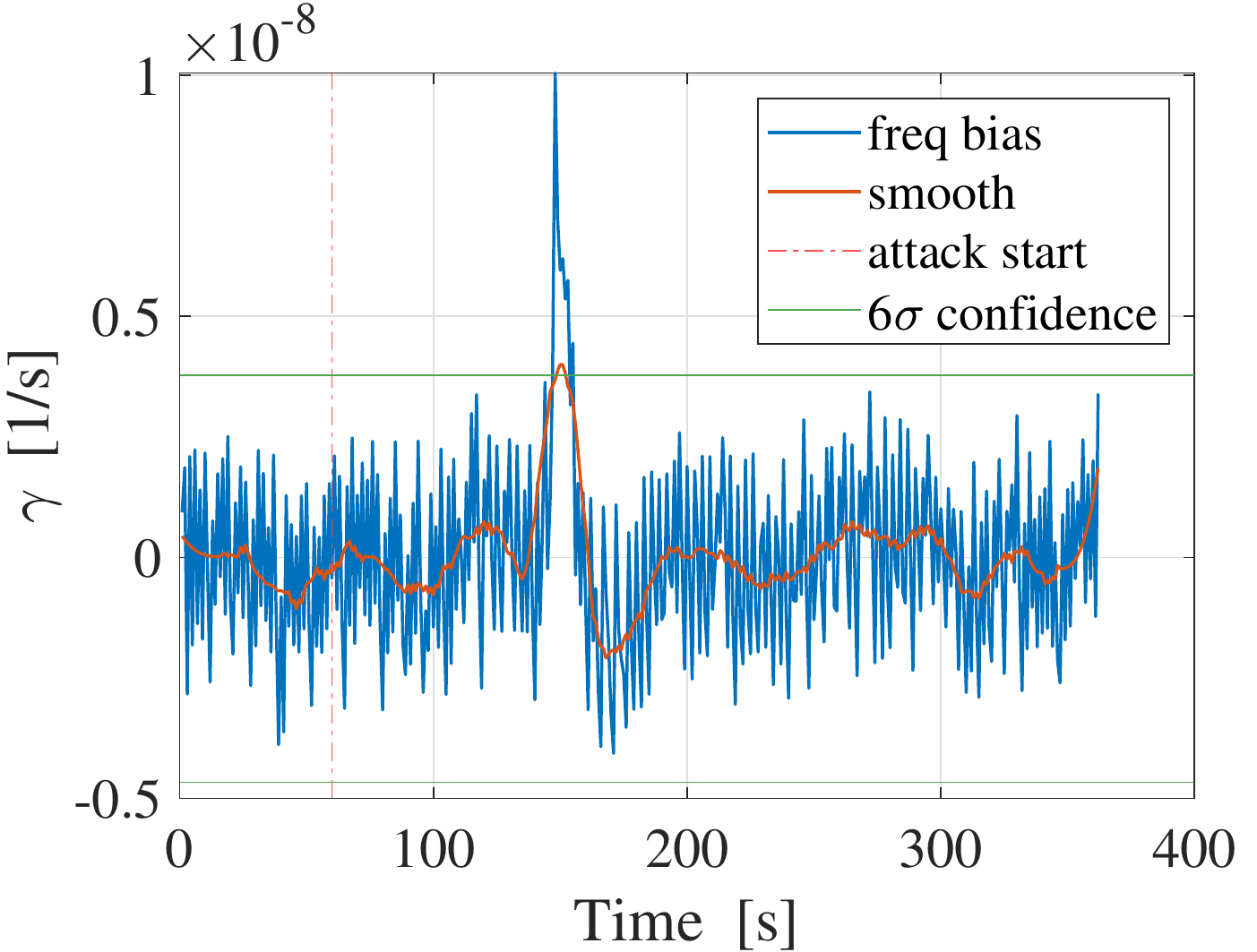}
    \caption{Frequency difference measured when under attack: the GNSS-disciplined clock frequency is subject to a sharp frequency variation at $T=150s$, which coincides with the adversarial lift-off.}
    \label{fig:freq-offset-difference-ref-gps-ds-3}
  \end{subfigure}
  \hfill%
  \begin{subfigure}[t]{0.33\textwidth}
    \includegraphics[width=\linewidth]{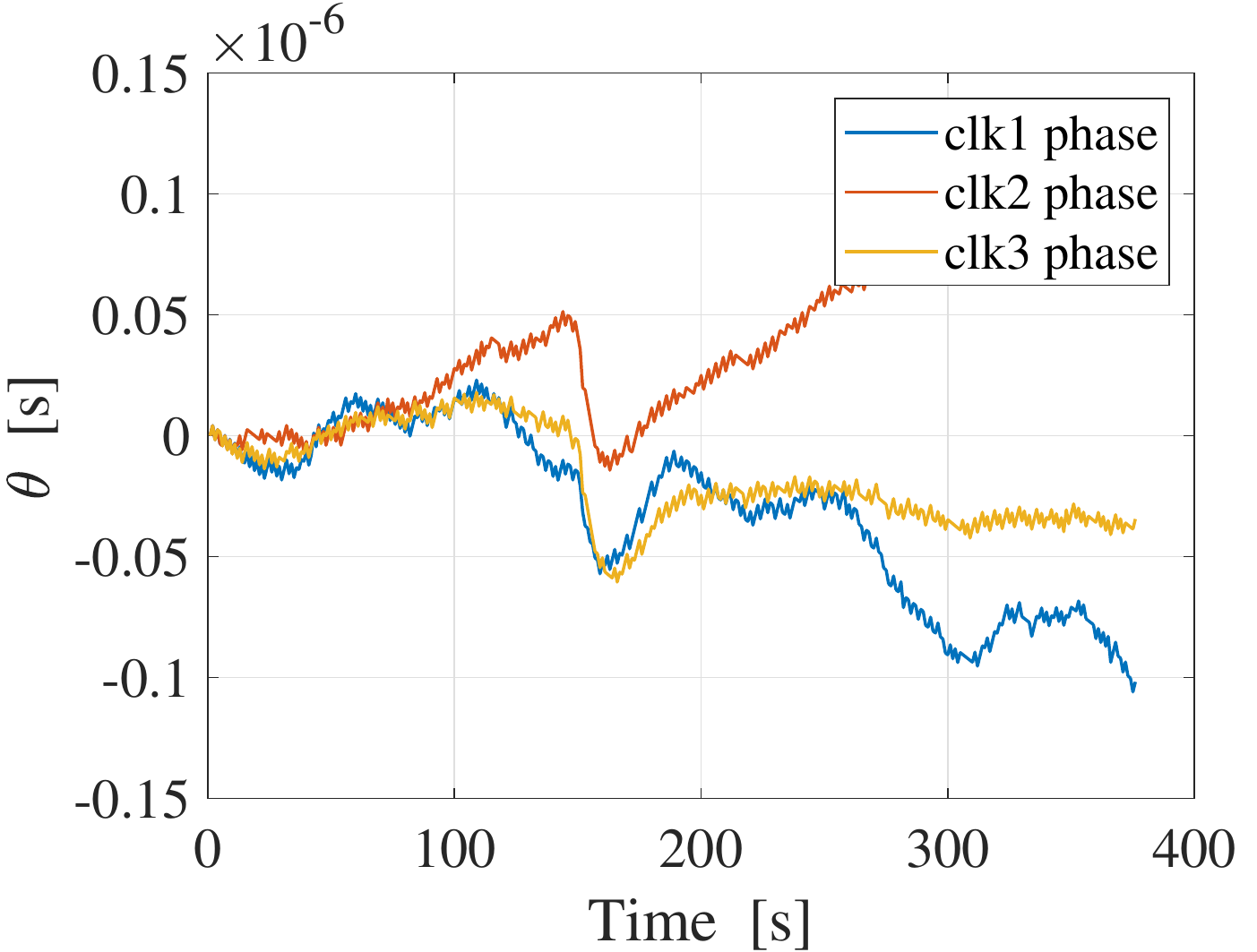}
    \caption{Phase difference correction measured with three independent oscillators. Correction values highlighting the discontinuity caused by the attacker at $T=150s$.}
    \label{fig:zero-mean-phases-ds-3}
  \end{subfigure}
  \caption{Static receiver and phased locked adversary (Scenario 3).}
  \label{fig:spoofed-ds-3}
\end{figure*}

In Scenario 3, the attacker behavior is more subtle and harder to detect, due to the phase lock with the real GNSS signals. Nevertheless, the beginning of the lift-off phase produces sharp discontinuities in the GNSS clock frequency that are successfully detected by the frequency offset test (Figure \ref{fig:freq-offset-difference-ref-gps-ds-3}). The discontinuity in the frequency state is more aggressive, but the overall shorter duration requires very precise measurements for detection. In this case, if the amplitude of the frequency change rate is beyond what specified by the manufacturer, an attack can be detected. 
On the other hand, the phase offset based test shows only small variations that are within the detector threshold (Figure \ref{fig:time-offset-difference-ref-gps-ds-3}). Investigations of the causes of this behavior at the receiver is ongoing.


In \cite{Humphreys2012}, a victim receiver is considered fully captured once the attacker achieves a \SI{2}{\micro \second} time offset induced in the victim. Notably, in Scenario 2 and 5, the total time offset at the detection threshold is below \SI{0.3}{\micro\second}, showing the ability of the proposed system to detect early the onset of an attack. Specifically, attacks against a static receiver are fully detected within \SI{40}{\second} (\SI{50}{\second} for mobile receivers) from the beginning of the victim manipulation (corresponding to the moment when the ensemble offset estimate crosses the $6\sigma$ confidence threshold). The output of the detector from \cref{fig:both-detector-output-ds-2,fig:both-detector-output-ds-5} show exactly this.
We achieved a detection latency more than twice lower compared to similar methods proposed in literature, with the combination of the frequency and phase tests. Additionally, the stability of the clock ensemble can be used to provide holdover in case of successful detection of an attack. This can be achieved leveraging the states estimated by the Kalman filter: once an attacker is fully detected, the GNSS offset and frequency information could be corrected based on the estimation of the other clocks in the ensemble, potentially providing accurate time for the duration of the attack (\cref{fig:zero-mean-phases-ds-2,fig:zero-mean-phases-ds-3,fig:zero-mean-phases-ds-5}). Further evaluations are ongoing, to characterize to which extent this is feasible and which level of accuracy in the holdver time solution is achievable.

\section{Conclusion}
\label{sec:conclusion}
We presented a constellation-agnostic, attack-agnostic detection method for advanced attackers tampering with the GNSS PVT by causing a modification of the time offset in the victim receiver. The method is based on a dual test on phase and frequency offsets, and allows low detection latency. For critical infrastructure applications, such as power grid monitoring, early detection can provide quick attack awareness, leaving more time to deploy countermeasures.  

Furthermore, the system is a standalone platform that can be integrated at the system level with the existing GNSS receiver, without requiring any modification to the receiver or the satellite infrastructure. This makes the enhanced receiver proposed in this work feasible for future integration in existing platforms. Additionally, it is not bound to specific clock characteristics and can be adapted depending on the application requirements: applications that are more sensitive to cost might trade detection performance. 

This method can work autonomously for extended periods of time when the receiver might be unable to retrieve network provided time and provide holdover once an attack is detected. On the other hand, complementing this solution with time obtained from a trusted network provider would extend the detection capabilities with correctness test of absolute time, allowing the receiver to thwart attackers active when the victim is in cold start. Additionally, the availability of precision local clocks reduces the number of time requests to the network infrastructure, as these two methods can cooperate to compensate the mutual shortcomings. 

Finally, it is important to notice that the proposed detection method based on precision clocks could be suitable for attacks against security enhanced GNSS signals. In case of attacks like SCER that target (civilian) authenticated GNSS signals, monitoring the receiver clock is important to detect replay and meaconing. Due to the complexity of setting up such a scenario, this is left for future work. 

\section*{acknowledgements}
This work was supported in part by the SSF SURPRISE cybersecurity project and the Security Link strategic research center.

\bibliographystyle{IEEEtran}
\bibliography{extracted}

\begin{thebibliography}{10}
\providecommand{\url}[1]{#1}
\csname url@samestyle\endcsname
\providecommand{\newblock}{\relax}
\providecommand{\bibinfo}[2]{#2}
\providecommand{\BIBentrySTDinterwordspacing}{\spaceskip=0pt\relax}
\providecommand{\BIBentryALTinterwordstretchfactor}{4}
\providecommand{\BIBentryALTinterwordspacing}{\spaceskip=\fontdimen2\font plus
\BIBentryALTinterwordstretchfactor\fontdimen3\font minus
  \fontdimen4\font\relax}
\providecommand{\BIBforeignlanguage}[2]{{%
\expandafter\ifx\csname l@#1\endcsname\relax
\typeout{** WARNING: IEEEtran.bst: No hyphenation pattern has been}%
\typeout{** loaded for the language `#1'. Using the pattern for}%
\typeout{** the default language instead.}%
\else
\language=\csname l@#1\endcsname
\fi
#2}}
\providecommand{\BIBdecl}{\relax}
\BIBdecl

\bibitem{walker2015galileo}
P.~Walker, V.~Rijmen, I.~Fern{\'{a}}ndez-Hern{\'{a}}ndez, L.~Bogaardt,
  G.~Seco-Granados, J.~Sim{\'{o}}n, D.~Calle, and O.~Pozzobon, ``{Galileo open
  service authentication: A complete service design and provision analysis},''
  in \emph{Proceedings of the 28th International Technical Meeting of the
  Satellite Division of The Institute of Navigation (ION GNSS+ 2015), Tampa,
  Florida}, vol.~5, 2015, pp. 3383--3396.

\bibitem{anderson2017chips}
J.~M. Anderson, K.~L. Carroll, N.~P. DeVilbiss, J.~T. Gillis, J.~C. Hinks,
  B.~W. O'Hanlon, J.~J. Rushanan, L.~Scott, and R.~A. Yazdi, ``{Chips-message
  robust authentication (chimera) for GPS civilian signals},'' in
  \emph{Proceedings of the 30th International Technical Meeting of the
  Satellite Division of The Institute of Navigation (ION GNSS+ 2017), Portland,
  Oregon}, vol.~4, 2017, pp. 2388--2416.

\bibitem{psiaki2016gnss}
M.~L. Psiaki and T.~E. Humphreys, ``{GNSS Spoofing and Detection},''
  \emph{Proceedings of the IEEE}, vol. 104, no.~6, pp. 1258--1270, 2016.

\bibitem{schmidtSurveyAnalysisGNSS2016}
D.~Schmidt, K.~Radke, S.~Camtepe, E.~Foo, and M.~Ren, ``{A survey and analysis
  of the GNSS spoofing threat and countermeasures},'' \emph{ACM Computing
  Surveys}, vol.~48, no.~4, pp. 1--31, 2016.

\bibitem{Humphreys2008}
T.~E. Humphreys, B.~M. Ledvina, M.~L. Psiaki, B.~W. O'Hanlon, and P.~M.
  Kintner, ``{Assessing the spoofing threat: Development of a portable gps
  civilian spoofer},'' in \emph{21st International Technical Meeting of the
  Satellite Division of the Institute of Navigation (ION GNSS 2008), Savanna,
  GA}, vol.~2, 2008, pp. 1198--1209.

\bibitem{borowski2012detecting}
H.~Borowski, O.~Isoz, F.~M. Eklof, S.~Lo, and D.~Akos, ``{Detecting false
  signals: With automatic gain control},'' \emph{GPS World}, vol.~23, no.~4,
  pp. 38--43, 2012.

\bibitem{tippenhauer2011requirements}
N.~O. Tippenhauer, C.~P{\"{o}}pper, K.~B. Rasmussen, and S.~{\v{C}}apkun, ``{On
  the requirements for successful GPS spoofing attacks},'' in \emph{Proceedings
  of the ACM Conference on Computer and Communications Security, New York, NY,
  USA}.\hskip 1em plus 0.5em minus 0.4em\relax ACM, 2011, pp. 75--85.

\bibitem{Humphreys2012a}
T.~Humphreys, J.~Bhatti, D.~Shepard, and K.~Wesson, ``{The Texas spoofing test
  battery: Toward a standard for evaluating GPS signal authentication
  techniques},'' \emph{Proceedings of the 25th International Technical Meeting
  of the Satellite Division of The Institute of Navigation (ION GNSS 2012),
  Nashville, TN}, vol.~5, pp. 3569--3583, 2012.

\bibitem{PapadimitratosJa:C:2008}
P.~Papadimitratos and A.~Jovanovic, ``{Protection and Fundamental Vulnerability
  of GNSS},'' in \emph{IEEE International Workshop on Satellite and Space
  Communications (IEEE IWSSC)}, Toulouse, France, October 2008, pp. 167--171.

\bibitem{papadimMilcom2008}
------, ``{GNSS-based positioning: Attacks and countermeasures},'' in
  \emph{Proceedings of the IEEE Military Communications Conference, San Diego,
  CA, USA}, 2008, pp. 1--71.

\bibitem{spangheroGNSS20}
M.~Spanghero, K.~Zhang, and P.~Papadimitratos, ``{Authenticated time for
  detecting GNSS attacks},'' in \emph{Proceedings of the 33rd International
  Technical Meeting of the Satellite Division of the Institute of Navigation,
  Virtual conference}, 10 2020, pp. 3826--3834.

\bibitem{kzmsppPLANS2020}
K.~Zhang, M.~Spanghero, and P.~Papadimitratos, ``{Protecting GNSS-based
  Services using Time Offset Validation},'' in \emph{Proceedings of the 2020
  IEEE/ION Position, Location and Navigation Symposium, Portland, OR, USA},
  2020, pp. 575--583.

\bibitem{Arafin2016DetectingOscillators}
M.~T. Arafin, D.~M. Anand, and G.~Qu, ``Detecting gnss spoofing using a network
  of hardware oscillators,'' in \emph{Proceedings of the Annual Precise Time
  and Time Interval Systems and Applications Meeting, Monterey, California}, 1
  2016, pp. 74--79.

\bibitem{Arafin2017}
M.~T. Arafin, D.~Anand, and G.~Qu, ``A low-cost gps spoofing detector design
  for internet of things (iot) applications,'' in \emph{Proceedings of the ACM
  Great Lakes Symposium on VLSI, Banff, Alberta, Canada}, 5 2017, pp. 161--166.

\bibitem{marnach2013detecting}
D.~Marnach, D.~Mauw, M.~Martins, and C.~Harpes, ``{Detecting Meaconing Attacks
  by Analysing the Clock Bias of GNSS Receivers},'' \emph{Artificial
  Satellites}, vol.~48, no.~2, pp. 63--83, 5 2013.

\bibitem{Hwang2014}
P.~Y. Hwang and G.~A. McGraw, ``Receiver autonomous signal authentication
  (rasa) based on clock stability analysis,'' \emph{IEEE Symposium on Position
  Location and Navigation, Monterey, CA, USA}, pp. 270--281, 2014.

\bibitem{Spanghero2021a}
M.~Spanghero and P.~Papadimitratos, ``{Detecting GNSS misbehaviour with
  high-precision clocks},'' in \emph{Proceedings of the 14th ACM Conference on
  Security and Privacy in Wireless and Mobile Networks, New York, NY,
  USA}.\hskip 1em plus 0.5em minus 0.4em\relax ACM, 6 2021, pp. 389--391.

\bibitem{Krawinkel2015}
T.~Krawinkel and S.~Sch{\"{o}}n, ``{Benefits of chip scale atomic clocks in
  GNSS applications},'' \emph{Proceedings of the 28th International Technical
  Meeting of the Satellite Division of The Institute of Navigation (ION GNSS+
  2015), Tampa, Florida}, vol.~4, pp. 2867--2874, 2015.

\bibitem{Gaggero2008}
P.~O. Gaggero and D.~Borio, ``{Ultra-stable oscillators: Limits of GNSS
  coherent integration},'' in \emph{Proceedings of the 21st International
  Technical Meeting of the Satellite Division of The Institute of Navigation
  (ION GNSS 2008), Savannah, GA}, vol.~1, 2008, pp. 101--111.

\bibitem{Levine1999c}
J.~Levine, ``Introduction to time and frequency metrology,'' \emph{Review of
  Scientific Instruments}, vol.~70, pp. 2567--2596, 06 1999.

\bibitem{shangClockSingleSignal}
S.~Shang, H.~Li, Y.~Wei, and M.~Lu, ``{GNSS spoofing detection and
  identification based on clock drift monitoring using only one signal},''
  \emph{Proceedings of the 2020 International Technical Meeting of The
  Institute of Navigation, San Diego, California}, pp. 331--341, 2020.

\bibitem{Shepard2012c}
\BIBentryALTinterwordspacing
D.~P. Shepard, T.~E. Humphreys, and A.~A. Fansler, ``{Evaluation of the
  vulnerability of phasor measurement units to GPS spoofing attacks},''
  \emph{International Journal of Critical Infrastructure Protection}, vol.~5,
  no. 3-4, pp. 146--153, 2012. [Online]. Available:
  \url{http://dx.doi.org/10.1016/j.ijcip.2012.09.003}
\BIBentrySTDinterwordspacing

\bibitem{JonesTryonTimeSeries}
R.~H. Jones and P.~V. Tryon, ``{Continuous Time Series Models for Unequally
  Spaced Data Applied to Modeling Atomic Clocks},'' \emph{SIAM Journal on
  Scientific and Statistical Computing}, vol.~8, no.~1, pp. 71--81, 1987.

\bibitem{Greenhall2001}
C.~A. Greenhall, ``{Kalman plus weights: a time scale algorithm},''
  \emph{Proceedings of the 33th Annual Precise Time and Time Interval Systems
  and Applications Meeting, Long Beach, California}, no.~11, pp. 445--454,
  2001.

\bibitem{Greenhall2011b}
C.~A. {Greenhall}, ``Reduced kalman filters for clock ensembles,'' in
  \emph{Proceedings of the 2011 Joint Conference of the IEEE International
  Frequency Control and the European Frequency and Time Forum (FCS), San
  Francisco, CA, USA}, 2011, pp. 1--5.

\bibitem{Zenzinger2012}
A.~Zenzinger, T.~Bartusch, C.~Kuehl, S.~Fischer, and A.~Shrestha, ``{Failure
  detection and correction for clock ensemble in space},'' in \emph{6th ESA
  Workshop on Satellite Navigation Technologies: Multi-GNSS Navigation
  Technologies Galileo's Here, NAVITEC 2012 and European Workshop on GNSS
  Signals and Signal Processing, Noordwijk, Netherlands}, 2012.

\bibitem{Peiffer2016}
B.~Peiffer, S.~Goguri, S.~Dasgupta, and R.~Mudumbai, ``{An approach to Kalman
  filtering for oscillator tracking},'' \emph{Proceedings of the 49th Asilomar
  Conference on Signals, Systems and Computers, Pacific Grove, CA, USA}, vol.
  2015-Febru, pp. 261--265, 2015.

\bibitem{USNOAccuracy}
\BIBentryALTinterwordspacing
U.S.\hspace{2pt}Space\hspace{2pt}Force, ``{GPS Accuracy}.'' [Online].
  Available: \url{http://www.gps.gov/systems/gps/performance/accuracy/}
\BIBentrySTDinterwordspacing

\bibitem{ESAAccuracy}
\BIBentryALTinterwordspacing
European\hspace{2pt}Space\hspace{2pt}Agency, ``{Galileo Accuracy}.'' [Online].
  Available:
  \url{http://gssc.esa.int/navipedia/index.php/Galileo\_Performances}
\BIBentrySTDinterwordspacing

\bibitem{Cantor1999}
S.~R. Cantor, T.~M. Corporation, B.~R. Bedford, A.~Stern, and H.~M. St,
  ``{Clock Technology},'' no. June.\hskip 1em plus 0.5em minus 0.4em\relax
  Proceedings of the 55th Annual Meeting of The Institute of Navigation,
  Cambridge, MA, 6 1999, pp. 28--30.

\bibitem{Kim2012}
H.~Kim, X.~Ma, and B.~R. Hamilton, ``{Tracking low-precision clocks with
  time-varying drifts using Kalman filtering},'' \emph{IEEE/ACM Transactions on
  Networking}, vol.~20, no.~1, pp. 257--270, 2012.

\bibitem{Kohno2005}
T.~Kohno, A.~Broido, and K.~C. Claffy, ``{Remote physical device
  fingerprinting},'' in \emph{IEEE Transactions on Dependable and Secure
  Computing}, vol.~2, no.~2, 4 2005, pp. 93--108.

\bibitem{Brown1991b}
K.~R. Brown, ``{The Theory of the GPS Composite Clock},'' \emph{Proceedings of
  the 4th International Technical Meeting of the Satellite Division of The
  Institute of Navigation (ION GPS 1991), Albuquerque, NM}, pp. 223--242, 1991.

\bibitem{Greenhall2006}
C.~A. Greenhall, ``{A Kalman filter clock ensemble algorithm that admits
  measurement noise},'' \emph{Metrologia}, vol.~43, no.~4, 2006.

\bibitem{ZEDF9PInterfaceDescription}
{UBlox}, ``{ZED-F9P Interface Description},'' 2020.

\bibitem{Blum2019}
R.~Blum, D.~D{\"{o}}tterb{\"{o}}ck, and T.~Pany, ``{Investigation of the
  vulnerability of mobile networks against spoofing attacks on their GNSS
  timing-receiver and developing a meaconing protection},'' \emph{Proceedings
  of the 2019 International Technical Meeting of The Institute of Navigation,
  Reston, Virginia}, pp. 345--362, 2019.

\bibitem{humphreys2013detection}
T.~E. Humphreys, ``{Detection strategy for cryptographic gnss anti-spoofing},''
  \emph{IEEE Transactions on Aerospace and Electronic Systems}, vol.~49, no.~2,
  pp. 1073--1090, 2013.

\bibitem{Hutsell1995Relating}
S.~Hutsell, ``{Relating the Hadamard Variance to MCS Kalman Filter Clock
  Estimation},'' \emph{Proceedings of the 27th Annual Precise Time and Time
  Interval Systems and Applications Meeting, San Diego, CA}.

\bibitem{razavi2000}
B.~Razavi, \emph{Design of Analog CMOS Integrated Circuits}, 1st~ed.\hskip 1em
  plus 0.5em minus 0.4em\relax USA: McGraw-Hill, Inc., 2000.

\bibitem{Humphreys2012}
T.~Humphreys, J.~Bhatti, D.~Shepard, and K.~Wesson, ``The texas spoofing test
  battery: Toward a standard for evaluating gps signal authentication
  techniques,'' pp. 3569--3583, 9 2012.

\end{thebibliography}

\thebiography
\begin{biographywithpic}
{Marco Spanghero}{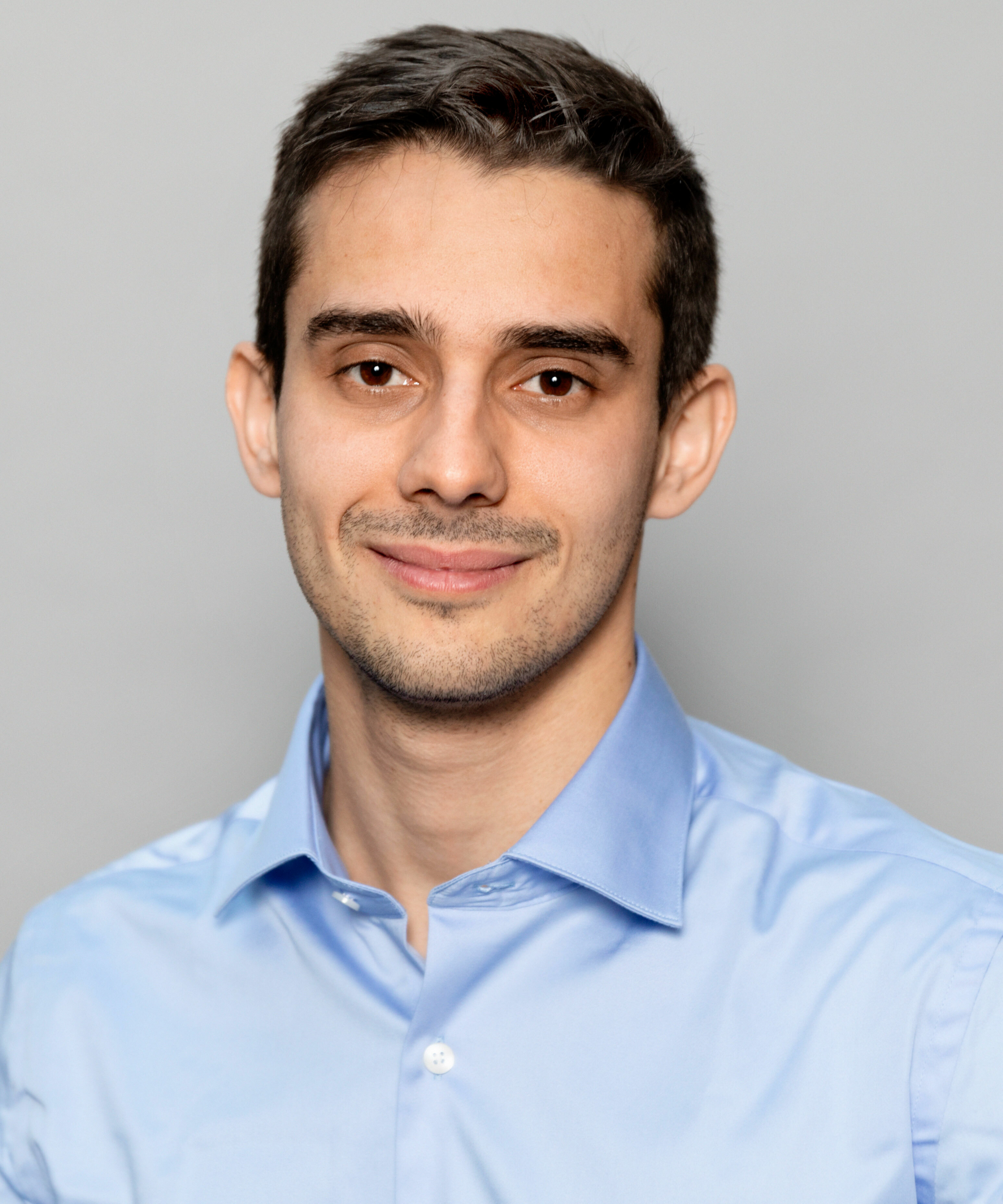}
received his B.S. in Electronics Engineering from Politecnico of Milano and MSc degree in Embedded System from KTH Royal Institute of Technology, Stockholm, Sweden. He is currently a Ph.D candidate with the Networked Systems Security (NSS) group at KTH Royal Institute of Technology, Stockholm, Sweden and associate to the WASP program from Knut and Alice Wallenberg foundation. His research interest is concerned with secure positioning and synchronization.
\end{biographywithpic} 

\begin{biographywithpic}
{Panagiotis 'Panos' Papadimitratos}{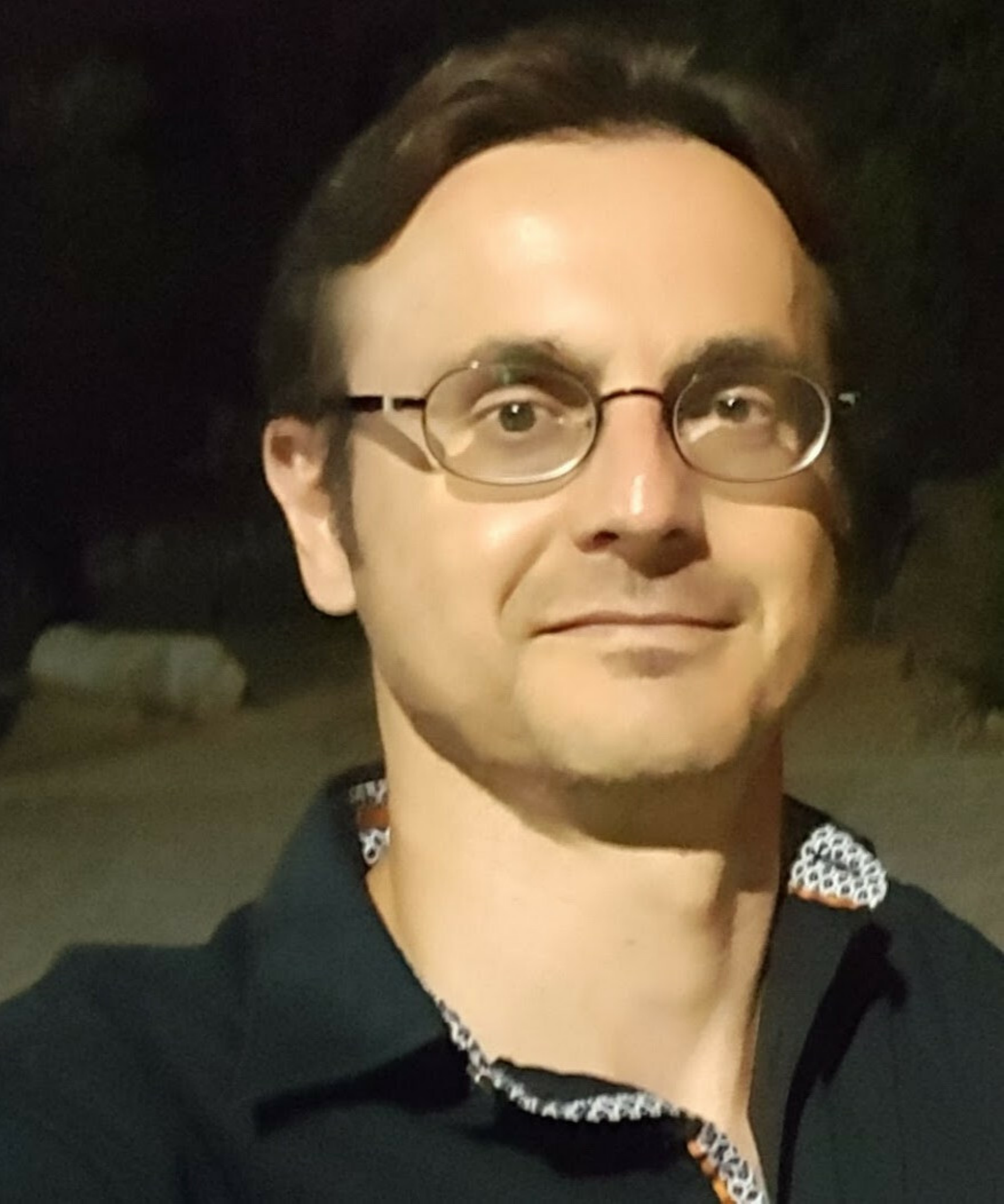}
(Fellow, IEEE) earned his Ph.D. degree from Cornell University, Ithaca, NY, USA. At KTH, Stockholm, Sweden, he leads the Networked Systems Security (NSS) group and he is a member of the Steering Committee of the Security Link Center. He serves or served as: member of the ACM WiSec and CANS conference steering committees and the PETS Editorial and Advisory Boards; Program Chair for the ACM WiSec’16, TRUST’16, and CANS’18 conferences; General Chair for the ACM WISec’18, PETS’19, and IEEE EuroS\&P’19 conferences; Associate Editor of the IEEE TMC, IEEE/ACM ToN and IET IFS journals, and Chair of the Caspar Bowden PET Award. Panos is a Fellow of the Young Academy of Europe, a Knut and Alice Wallenberg Academy Fellow, and an ACM Distinguished Member. NSS webpage: \url{https://www.eecs.kth.se/nss}.
\end{biographywithpic}

\end{document}